\documentclass{aa}  
\usepackage{graphicx}
\usepackage{txfonts}
\usepackage{natbib}
\begin{document}
   \title{Galactic abundance gradients from Cepheids:\thanks{Based on observations obtained at the Canada-France-Hawaii
Telescope (CFHT), which is operated by the National Research Council
of Canada, the Institut National des Sciences de l'Univers of the Centre
National de la Recherche Scientifique of France, and the University of
Hawaii. Based on observations collected with FEROS at the European
Southern Observatory (La Silla, Chile) under proposal ID: 60.A-9120. Based on observations obtained
at the Telescope Bernard Lyot (USR5026) operated by the Observatoire
Midi-Pyr\'en\'ees and the Institut National des Science de l’Univers of the
Centre National de la Recherche Scientifique of France. The data of NARVAL were reduced using the 
data reduction software Libre-ESpRIT, written and provided by J.-F. Donati from the LATT 
(Observatoire Midi-Pyr\'en\'ees)}}

   \subtitle{$\alpha$ and heavy elements in the outer disk.}

   \author{B.~Lemasle
          \inst{1,2}
          \and
          P.~Fran\c cois\inst{3,4}
	  \and
          K.~Genovali\inst{5}
	  \and
	  V.~V.~Kovtyukh\inst{6}
	  \and
	  G.~Bono\inst{5,7}
	  \and
	  L.~Inno\inst{5,10}
	  \and
	  C.~D.~Laney\inst{8,9}
	  \and
	  L.~Kaper\inst{1}
	  \and
	  \newline M. Bergemann\inst{10}
	  \and
	  M. Fabrizio\inst{11}
	  \and
	  N. Matsunaga\inst{12}
	  \and
	  S.~Pedicelli\inst{5}
	  \and
	  F.~Primas\inst{13}
	  \and
	  M.~Romaniello\inst{13}          
}

   \institute{Astronomical Institute ‘Anton Pannekoek’, Science Park 904, P.O. Box 94249, 1090 GE Amsterdam, The Netherlands \\B.J.P.Lemasle@uva.nl
	 \and         
	      Kapteyn Astronomical Institute, Landleven 12, 9747 AD Groningen, The Netherlands 
	 \and
	      GEPI - Observatoire de Paris, 64 Avenue de l'Observatoire, 75014 Paris, France
	 \and
	      UPJV - Universit\'e de Picardie Jules Verne, 80000 Amiens, France
	 \and
              Dipartimento di Fisica, Universit\`a di Roma Tor Vergata, via della Ricerca Scientifica 1, 00133 Rome, Italy
	 \and
	      Astronomical Observatory of Odessa National University, and Isaac Newton Institute of Chile, Odessa branch, Shevchenko Park, 65014, Odessa, Ukraine
	 \and
	      INAF - Osservatorio Astronomico di Roma, via Frascati 33, Monte Porzio Catone, Rome, Italy
	 \and
	      South African Astronomical Observatory, P.O. Box 9, Observatory 7935, South Africa
	 \and
	      Department of Physics and Astronomy, Western Kentucky University, 1906 College Heights Blvd, Bowling Green, KY 42101-1077,USA
	 \and
	      Max Planck Institute for Astrophysics, Karl-Schwarzschild-Str. 1, D-85741 Garching bei Munchen, Germany
	 \and
	      INAF - Osservatorio Astronomico Collurania, Via M. Maggini, 64100 Teramo, Italy
	 \and
	      Department of Astronomy, School of Science, The University of Tokyo, 7-3-1 Hongo, Bunkyo-ku, Tokyo 113-0033, Japan
	 \and
	      European Southern Observatory, Karl-Schwarzschild-Str. 2, D-85748 Garching bei Munchen, Germany 
             }

   \date{Received September 15, 1996; accepted March 16, 1997}


  \abstract
   {Galactic abundance gradients set strong constraints to chemo-dynamical evolutionary models of the Milky Way. Given the period-luminosity relations that provide accurate distances and the large number of spectral lines, Cepheids are excellent tracers of the present-day abundance gradients.}
   {We want to measure the Galactic abundance gradient of several chemical elements. While the slope of the Cepheid iron gradient did not vary much from the very first studies, the gradients of the other elements are not that well constrained. In this paper we focus on the inner and outer regions of the Galactic thin disk.} 
   {We use high-resolution spectra (FEROS, ESPADONS, NARVAL) to measure the abundances of several light (Na, Al), $\alpha$~(Mg, Si, S, Ca), and heavy elements (Y, Zr, La, Ce, Nd, Eu) in a sample of 65 Milky Way Cepheids. Combining these results with accurate distances from period-Wesenheit relations in the near infrared enables us to determine the abundance gradients in the Milky Way.}
   {Our results are in good agreement with previous studies on either Cepheids or other tracers. In particular, we confirm an upward shift of $\approx$0.2 dex for the Mg abundances, as has recently been reported. We also confirm the existence of a gradient for all the heavy elements studied in the context of a local thermodynamic equilibrium analysis. However, for Y, Nd, and especially La, we find lower abundances for Cepheids in the outer disk than reported in previous studies, leading to steeper gradients. This effect can be explained by the differences in the line lists used by different groups.}
   {Our data do not support a flattening of the gradients in the outer disk, in agreement with recent Cepheid studies and chemo-dynamical simulations. This is in contrast to the open cluster observations but remains compatible with a picture where the transition zone between the inner disk and the outer disk would move outward with time.}
   

   \maketitle
%

\section{Introduction}

\par The distribution of the elements along the disk of galaxies is influenced by the star formation history (SFH), accretion events, radial flows of gas, and radial mixing of stars. It usually follows a radial gradient. Radial abundance gradients are therefore used to constrain models describing the formation and evolution of galaxies, in particular of the Milky Way.\\

\par Different observational tracers have been used to derive the Galactic abundance gradients: HII regions, O/B-type stars, Cepheids, open clusters and planetary nebulae (PNe). Because the nature of these objects is different, it allows us to trace different elements and age groups, paving the way for time evolution studies that are still in their first faltering steps.
\par Compared to the other tracers, Cepheids present several advantages: i) as famous primary distance indicators, they provide very accurate distances; ii) they are bright stars and can thus still be observed at large distances from the Sun; iii) they are cool supergiants that present a large set of well-defined absorption lines. Therefore accurate abundances of many elements can be determined. On the other hand, the population of Galactic Cepheids is still relatively poorly sampled, and the young age of these stars \citep[$<$200 Myr,][]{Bono2005} allows us to study only the present-day gradient. As Cepheids are pulsating variable stars, one could wonder whether the phase of observation would influence the abundance determination. However,  \citet{Luck2004}, \citet{Kov2005a}, \citet{And2005}, and \citet{Luck2008} have demonstrated in a series of papers that results are extremely stable throughout the period, whatever the period of the star.  
\par Since the early works of \cite{Har1981,Har1984}, the determinations of the Galactic iron gradient from Cepheids have been remarkably consistent: \citet{And2002a,And2002b,And2002c}, \citet{Luck2003}, \citet{And2004}, \citet{Kov2005b}, \citet{Luck2006,Luck2011a}, \citet{Luck2011b}, \citet{Lem2007,Lem2008}, and \citet{Pedi2009,Pedi2010} all report a value close to --0.06 dex/kpc. Cepheid abundances have also been derived by \citet{Yong2006,Us2011a,Us2011b}.\\

\par These values are in good agreement with studies based on other tracers, even if a direct comparison between them is quite difficult: as already mentioned, different objects trace different elements and different ages. The abundances in HII regions can be derived from radio recombination lines, collisionally excited lines, or optical recombination lines that lead to somewhat discrepant abundances (the same also holds for PNe). Therefore the [O/H] gradients span quite a large range of values, which is between $-0.030$ dex/kpc and --0.08 dex/kpc if we consider only the recent determinations \citep[e.g.,][]{Sim1995,Aff1996,Aff1997,Deh2000,Qui2006b,Rud2006,Balser2011}.
\par The gradients obtained with O-B1 stars vary between --0.03 and --0.07 dex/kpc \citep[e.g.,][]{Sma1997,Gum1998,Daf2004b,Roll2000}. As O-B1 stars are younger than 10 Myr, they trace the present-day gradient even more closely than Cepheids.
\par There are numerous abundance studies of open clusters from different groups: \citet{Bra2008}, \citet{Carra2007}, \citet{Chen2003}, \citet{Frie2002,Frie2010}, \citet{Jaco2008,Jaco2009,Jaco2011a,Jaco2011b}, \citet{Mag2009a,Mag2010}, \citet{Pan2010}, \citet{Ses2006,Ses2007}, \citet{Yong2005,Yong2012}. They all reach the same conclusion: a linear gradient of approximately --0.06 dex/kpc extending quite far into the outer disk, and a flattening at a level of [Fe/H] $\approx$ --0.3 dex, somewhere between 10 and 14 kpc. The only exception is \citet{Ses2008}, who report a larger slope of --0.17 dex/kpc in the inner part of the disk.
\par Radial [O/H] gradients from PNe have values generally comprised between --0.02 dex/kpc and~--0.06 dex/kpc \citep{Mac1999,Cos2004,Hen2004,Perr2006,Hen2010}. \citet{Pott2006} obtain better accuracies for their abundances because they measure transitions both in the visible and in the infrared. However, their sample does not extend beyond $\approx$10 kpc (where the most metal-poor PNe are found), which might be why they report a stronger gradient of --0.085 dex/kpc. In contrast to the other studies, \citet{Stan2006} find no evidence of a gradient.
\par The time evolution of the metallicity gradients is still very controversial: from PNe, \citet{Mac2003} found a flattening of the gradient with time, while \citet{Stan2010} report instead a steepening of the gradient and \citet{Hen2010} found no evolution with time. Most studies dividing open clusters into age bins report a flattening with time. When combining several tracers, \citet{Mac2009} again found a flattening with time, while \citet{Mag2009a} report only marginal evidence of gradients flattening with time in the 7-12 kpc range. From a comparison between Open Clusters and Cepheids, \citet{Yong2012} report a flattening of the gradient with time in the [5--12] kpc range for all the elements considered in their study.\\ 

\par However, a simple, linear gradient may not be the better way to describe the distribution of the elements in the Galactic disk. From Cepheids, \citet{And2002b}, \citet{Pedi2009,Pedi2010}, and \citet{Geno2013a} found a steeper gradient in the very inner disk ($\leq$~7~kpc). We have already mentioned that most of the studies based on open clusters find a flatter gradient in the outer disk. In this region, the situation is much less clear when considering other objects: e.g., \citet{Vil1996} (HII regions) and \citet{Cos2004} (PNe) report flat gradients in the outskirts of the Milky Way, while other authors using the same tracers do not. From Cepheids, a flattening gradient seemed well established \citep{And2002c,Luck2003,And2004,Lem2008}, but the recent addition of outer disk Cepheids in the sample \citep{Luck2011a,Luck2011b} weakens this hypothesis. A flattening gradient is also seen in external galaxies, \citep[e.g.,][]{Bre2009a}. Moreover, \citet{Yong2006} brought forward an interesting feature: the flat structure in the outer disk may occur with two different basement values, one at --0.5 dex and a second one as low as --0.8 dex, possibly related to a merger event.
\par In addition, \citet{Twa1997} suggested that instead of a break in the slope the radial metallicity gradient could experience an abrupt drop of the order of 0.2--0.3 dex around 10 kpc. This feature was also proposed in other studies \citep[e.g.,][]{Daf2004b,And2004}. From the open cluster point of view, it is still not clear whether this feature is real or only due to a too coarse sampling, despite the rapidly growing number of studies investigating this zone. Whether the metallicity distribution takes the shape of a step or more simply of a change in the slope, the exact location of the transition could be a bit further out in the disk and depend on the cluster ages \citep[e.g.,][]{Jaco2011b}. For Cepheids, the step was probably an artifact due to a sampling effect \citep{Lem2008,Luck2011a,Luck2011b}. Such a feature in the outer disk is generally supported by theoretical works invoking variations of the gas density, the presence of a bar, or the consequence of long-lived spiral structures \citep{Sca2013}.\\

\par Negative abundance gradients, comparable to those of the Milky Way, have also been reported in external spiral galaxies: e.g., M31 \citep{San2010}, M33 \citep{Bar2007,Mag2009b}, M83 \citep{Bre2009a}, NGC 300 \citep{Ku2008,Vla2009,Bre2009b}, NGC 7793 \citep{Vla2011}. In most cases the gradient also flattens in the outer disk.\\

\par The chemical enrichment of the Galaxy was first studied by pure chemical evolution models, most of them based on the so-called {\it inside-out} scenario \citep[e.g.,][]{Matt1989,Chia1997,Pran2000,Chia2001,Molla2005,Fu2009} in which the Milky Way primarily forms during two episodes of gas infall: the first one giving birth to the halo and the bulge and the second one producing the disk. During the second infall, the pristine gas accumulates more slowly and preferentially in the inner parts of the disk, leading to a faster chemical enrichment of these regions. Therefore a negative abundance gradient naturally arises within the Galactic disk. In particular, the model from \cite{Ces2007} computes radial gradients for numerous elements and closely reproduces Cepheid-based observations. Radial flows of gas \citep[up to a few km/s, see ][]{Lacey1985,Port2000,Spi2011} must also be taken into account to properly fit the observational data, in particular to recover the gas density distribution. Radial gas flows also enable steep enough gradients in the models (\citet{Spi2011}, but see \citet{Fu2013}) to be obtained, although the alternative explanation of a variable efficiency in the star formation rate could have a similar effect \citep[][and references therein]{Cola2008}.\\

\par More recently, dynamical aspects have also been taken into account through N-body/smoothed particle hydrodynamics (SPH), semi-analytical models, or cosmological simulations of galactic disks \citep[e.g.,][for the latest ones]{Stin2010,Rahi2011,Koba2011,Few2012}. \cite{Wier2011} systematically investigated the impact on the outcome when, not specifically looking for Milky Way analogs, they varied a large number of physical processes. On the other hand, \cite{Pil2012} compared different sets of simulations and two classical chemical evolution models, all fine-tuned to resemble the Milky Way. All of the models were able to reproduce the present-day gradient, but following different paths.
\par The radial migration of stars along a Galactic radius has recently been recognized as a key process to understanding the evolution of gradients with time. Several mechanisms have been proposed to explain the radial mixing: 
\begin{itemize}
	\item the position of stars around their birth radii shows an increased spread due to scattering by molecular clouds, which increases their epicyclic energy \citep[``blurring'',][]{Spi1953,Bin2008};
	\item the orbit of stars is moved inwards or outwards either by resonant scattering by the transient spiral structure \citep{Sell2002,Ros2008a,Ros2008b} or by the resonance overlap of the bar and spiral structure \citep{Schon2009,Min2010}. This mechanism is referred to as ``churning'' and implies a change in the angular momentum of the star;
	\item radial mixing can also be caused by the perturbation induced by a low-mass orbiting satellite \citep{Quil2009,Bird2012} or a satellite bombardment \citep{Bird2012}.
\end{itemize}
There were early models combining dynamics and chemical evolution \citep[e.g.,][]{Fran1993}. \cite{Schon2009} were the first to incorporate the recent results on radial migration of stars in chemical evolution models. Very recently, \cite{Min2012} presented new Milky Way models, where the chemical evolution is merged with disk numerical simulations. Each particle in their simulations represents only one star, while the N-body simulations with the highest resolution and an improved treatment of the chemistry \citep[e.g.,][]{Bird2012b} have particles of $\approx$ 10$^{4}$ M$_{\sun}$. Among other results, they found that radial migration affects the old stars more strongly. Radial migration in the disk is at the origin of azimuthal variations in the metallicity distribution of its old stars \citep{diMatt2013}\\

\par In this paper we investigate the shape of the gradient for 11 elements and focus on the outer disk. The paper is organized as follows: in Sect. \ref{Method}, we briefly recall the observations and data analysis, and in Sect. \ref{Dist} we describe the distances of the Cepheids based on near-infrared (NIR) photometry. Results are presented and discussed in Sect. \ref{Abund} (abundances) and Sect. \ref{Res} (gradients). Sect. \ref{Con} summarizes our findings.

\section{Observations and Method}
\label{Method}

\subsection{Data}

Our study is based on the same high-resolution spectra (ESPaDOnS: 25 stars, FEROS: 40 stars, Narval@TBL: 1 star) as in \cite{Lem2007,Lem2008} and \citet{Pedi2010}, where we determined the atmospheric parameters and iron abundances in order to study the iron gradient across the Galactic disk. The signal to noise (S/N) ratio ranges from 40 to 140, with most of the spectra having S/N$\simeq$100. 

\subsection{Atmospheric parameters}

We used the atmospheric parameters previously derived in the papers quoted above. For this reason, we only briefly summarize the method here. Because Cepheids are pulsating variable stars, it is very unlikely to obtain simultaneous photometric measurements. Therefore the atmospheric parameters have to be determined exclusively from spectroscopic information. To accurately determine $T_{\rm eff}$, which is critical in the process of abundance determination, we used the line-depth ratios method described in \cite{KovGor2000} and \cite{Kov2007}. This method has been long proven to give accurate temperatures and can also be double-checked because lines with both high and low $\chi_{\rm ex}$ values have to properly fit the curve of growth. The surface gravity, $\log g$, and the microturbulent velocity, $v_{\rm t}$, were determined by a classical spectroscopic analysis: We imposed i) the ionization balance between Fe I and Fe II with the help of the curve of growth and ii) that the Fe I abundance depends neither on line strength nor on the excitation potential. It should be noted here that Andrievsky and collaborators use a slightly modified version of the canonical analysis in order to avoid a possible influence of non local thermodynamic equilibrium (NLTE) effects on Fe I lines \citep{Kov1999}: surface gravities and microturbulent velocities are derived from Fe II lines instead of Fe I lines. However, the two methods lead to differences that are typically equal to or less than 0.10 dex for the [Fe/H] abundances.

\subsection{Line list}

Some usual lines in stellar spectroscopic analysis cannot be used to determine the chemical composition of Cepheids as they are much too wide in supergiants. For instance, the Mg I line at 5528.41 \AA{} as well as the Ba II lines at 6141.73 and 6496.91\AA{} have equivalent widths (EW) larger than 250 m\AA{} in most of the stars of our sample. This is the reason why the first study of barium in Cepheids (which takes into account NLTE effects) appeared only very recently \citep{And2013}. Surprisingly, it is also the first large study of the barium abundance across the Galactic disk. Moreover, Cepheids span a very wide range of $T_{\rm eff}$, $\log g$, and $v_{\rm t}$ along the pulsation cycle. In particular, $T_{\rm eff}$ can span a range larger than 1000 K. As abundances should ideally be derived from the same set of lines for the whole sample, one wants to discard lines that would disappear or become too strong at the edges of the $T_{\rm eff}$ range. The same outcome applies to the other atmospheric parameters. Finally, in the phases with the lowest $T_{\rm eff}$ and $\log g$, some parts of the spectrum, especially toward the blue, are so crowded that line blending makes it almost impossible to locate the continuum, preventing us from measuring lines in those parts of the spectrum.

\par In order to build a proper line list, we selected a dozen of our
spectra with a good S/N that covered the atmospheric parameter space 
(including [Fe/H]) as much as possible. From those spectra, we
picked up the lines fulfilling all the aforementioned criteria, namely,
lines that are always present but never too strong, with no or limited
blending, and with a continuum well defined around the line. Despite this
careful selection, it could still happen that some of the lines were
not measurable, either because the Cepheids were observed in a phase
that was less favorable or more simply because we could only reach a
limited S/N. For all the lines, we adopted the physical properties
(oscillator strength, excitation potential) listed in the Vienna
Atomic Line Database (VALD) \citep{Kup1999,Ryab1999}. Our line list
for the $\alpha$ and heavy elements can be found in
Table~\ref{linelist}. The Fe I and Fe II lines that were used to
determine the atmospheric parameters of the Cepheids are listed in \citet{Rom2008} and \citet{Pedi2010}; the EW of iron lines were not remeasured here. 

\subsection{Abundance determinations}

The atmospheric parameters are used as input for the MARCS models of \cite{Edv1993}. In this version of the MARCS models, the stellar atmosphere is described assuming a plane-parallel stratification, hydrostatic equilibrium, and local thermodynamic equilibrium (LTE). The EW of $\alpha$ and heavy elements were measured with a semi-interactive code ({\it fitline}, written by P. Fran\c cois) based on genetic algorithms from \citet{Char1995} (see \cite{Lem2007} for details). We adopted the solar chemical abundances provided by \citet{Gre1996}. For a given element, the final abundance of a star is estimated as the mean value of the abundances determined for each line of this element.\\
\indent In order to test how the uncertainties on the atmospheric parameters of the star affect the final abundance results, we computed the abundances with over- or underestimated values of T$_{eff}$ ($\pm$100~K), $log~g$ ($\pm$0.3~dex), v$_{t}$ ($\pm$0.5~dex), and [Fe/H] ($\pm$0.12~dex). The sum in quadrature of these uncertainties on the atmospheric parameters gives uncertainties on the abundances that are listed in Table~\ref{error} (uncertainties of the order of 0.12~dex on [Fe/H] \citep{Lem2008} leave the abundances unchanged and are therefore not included in Table~\ref{error}). We repeated the exercise for two stars spanning the whole temperature range for Cepheids: SX Vel (T$_{eff}$=6248~K) and SV Mon (T$_{eff}$=4900~K) 
\begin{table}[!ht]
\centering
\begin{tabular}{r|cccc}
\hline\hline
         & $\Delta$T$_{eff}$ & $\Delta$$log~g$ & $\Delta$v$_{t}$ & $\Delta$[X/H] \\
 Element &    (--100 K)      &    (--0.3~dex)  &    (+0.5~dex)   &               \\
         &         dex       &        dex      &         dex     &         dex   \\
\hline
 {[}Al/H] & --0.04 &  +0.02 & --0.02 & 0.05 \\
 {[}Ca/H] & --0.07 &  +0.03 & --0.09 & 0.12 \\
 {[}Ce/H] & --0.07 & --0.08 & --0.13 & 0.17 \\
 {[}Eu/H] & --0.06 & --0.08 & --0.04 & 0.11 \\
 {[}La/H] & --0.07 & --0.07 & --0.03 & 0.11 \\
 {[}Mg/H] & --0.06 &  +0.02 & --0.08 & 0.11 \\
 {[}Na/H] & --0.06 &  +0.02 & --0.16 & 0.17 \\
 {[}Nd/H] & --0.09 & --0.06 & --0.04 & 0.12 \\
 {[}S/H]  &  +0.00 & --0.04 & --0.11 & 0.12 \\
 {[}Si/H] & --0.06 &  +0.02 & --0.07 & 0.10 \\
 {[}Y/H]  & --0.05 & --0.09 & --0.05 & 0.12 \\	 
 {[}Zr/H] & --0.04 & --0.09 & --0.06 & 0.12 \\
\hline\hline
         & $\Delta$T$_{eff}$ & $\Delta$$log~g$ & $\Delta$v$_{t}$ & $\Delta$[X/H] \\
 Element &    (+100 K)       &    (+0.3~dex)   &   (--0.5~dex)   &               \\
         &         dex       &        dex      &         dex     &         dex   \\
\hline
 {[}Al/H] &  +0.06 & --0.03 &  +0.09 & 0.12 \\
 {[}Ca/H] &  +0.07 & --0.02 &  +0.03 & 0.08 \\
 {[}Ce/H] &  +0.00 &  +0.13 &  +0.04 & 0.14 \\
 {[}Eu/H] & --0.01 &  +0.13 &  +0.10 & 0.17 \\
 {[}La/H] &  +0.02 &  +0.13 &  +0.14 & 0.19 \\
 {[}Mg/H] &  +0.05 & --0.02 &  +0.13 & 0.14 \\
 {[}Na/H] &    -   &    -   &    -   &  -   \\
 {[}Nd/H] &  +0.00 &  +0.13 &  +0.07 & 0.15 \\
 {[}S/H]  &    -   &    -   &    -   &  -   \\
 {[}Si/H] &  +0.05 &  +0.02 &  +0.09 & 0.11 \\
 {[}Y/H]  & --0.01 &  +0.14 &  +0.13 & 0.19 \\	 
 {[}Zr/H] & --0.01 &  +0.12 &  +0.20 & 0.24 \\
\hline
\end{tabular}
\caption{{\it Top:} Error budget for SX Vel (T$_{eff}$=6250 K, $log~g$=1.3 dex, v$_{t}$=2.8 km/s, [Fe/H]=--0.2 dex). {\it Bottom:} Error budget for SV Mon (T$_{eff}$=4900 K, $log~g$=0.50 dex, v$_{t}$=3.4 km/s, [Fe/H]=--0.1 dex).}
\label{error}
\end{table}

\section{Distances}
\label{Dist}

\subsection{Period-Wesenheit relations}

For a given star, the Wesenheit index \citep[e.g.,][]{Mado1982} is a linear combination of a selected magnitude and one of its related colors. By construction, the Wesenheit index is a reddening-free quantity. For instance, the Wesenheit index for J and K magnitudes W$_{JK}$ is defined as
\begin{equation}
W_{JK} = K - \frac{A_{K}}{E(J-K)} \times (J-K).
\end{equation}
Compared to traditional period-luminosity(-color) relations, the period-Wesenheit (P-W) relations present several advantages.
\begin{itemize}
   \item They are almost linear over the entire period range;
   \item the slopes of P-W relations are almost independent of the metallicity;
   \item the dispersion of the P-W relations is significantly reduced compared to their corresponding period-luminosity (P-L) relations;
   \item P-W relations are reddening-free, whereas the use of classical P-L relations requires the determination of the individual E(B-V) for each Cepheid.
\end{itemize}
These characteristics of P-W relations have been predicted by theoretical pulsation models \citep[e.g.,][]{Fio2007,Bono2008}. They have been supported from an observational point of view, in particular as far as the P-W$_{VI}$ is concerned, because a very large amount of data is already available in those bands \citep[e.g.,][and references therein]{Fou2007,Mado2009,Bono2010,Ngeow2012}. The metallicity dependence of the optical P-W$_{VI}$ relations remains, however, controversial (\cite{Storm2011a,Storm2011b}, and \cite{Sha2011,Ger2011}, but see also \cite{Maja2011} for a discussion).

\subsection{Method}

New NIR P-W relations for Galactic Cepheids have been determined by \cite{Inno2013} for the J, H, and K bands. Their slopes have been calibrated with Cepheids in the Large Magellanic Cloud and their zero-points with parallax distances from the Hubble Space Telescope available for Galactic Cepheids. In order to reduce the sources of uncertainties, first overtone (FO) pulsators have not been fundamentalized. Instead, different sets of P-W relations for either fundamental or FO pulsators have been calibrated.

We adopt here the individual distances estimated by \cite{Geno2013b} using these new P-W relations and single-epoch measurements or preferably mean magnitudes in the NIR bands. The NIR magnitudes used in \cite{Geno2013b} originate from four different data samples, ordered below according to their photometric accuracy and light curve coverage:
\begin{enumerate} 
  \item NIR photometry of 132 northern Cepheids from \cite{Mon2011} with a complete coverage of the light curves;
  \item data from the South African Astronomical Observatory previously published by \cite{Laney2007} and references therein, with a complete coverage of the light curves;
  \item new, unpublished observations kindly provided by C. D. Laney (priv. comm.); the coverage of the light curves is currently only partial (6-12 points);   
  \item the 2MASS catalog, with only single-epoch observations. The mean magnitude is then estimated with a template light curve, the amplitude in the V-band and the epoch of maximum both available in the literature \citep{Sos2005}.
\end{enumerate}
Three values of the distance modulus (and relative errors) have been computed in the three NIR bands, the conclusive value being their weighted mean. The results are in very good agreement with previous results from \cite{Mon2011} and \cite{Storm2011a,Storm2011b}, in particular because they show no spurious trends with period or reddening. The reader interested in a complete description of the method adopted to estimate accurate distances together with a detailed analysis of the error budget is referred to \cite{Inno2013} and \cite{Geno2013b}. However, it is worth mentioning that the error budget includes the errors affecting the Galactocentric distance of the Sun. Table \ref{dist} lists the Galactocentric distances for our high-resolution sample. Our whole sample \citep[see][]{Geno2013b} is, to date, the largest homogeneous sample for accurate NIR distances of Galactic classical Cepheids.\\
    
\par We adopted 7.94 kpc as the Galactocentric distance of the Sun, derived from type II Cepheids and RR Lyrae stars by \cite{Groe2008}. \cite{Fritz2011} obtained the same value from Red Clump stars in the Galactic center (GC) and an updated extinction curve toward the GC. A very similar value (7.9 kpc) is derived by \cite{Reid2009} from the trigonometric parallax of Sgr B2, a massive star forming region very close to the GC. Similar values are reported by \cite{Matsu2013} from short-period variable stars close to the GC, by \cite{Trip2008} from the statistical parallax of late-type stars in the GC (8.07 kpc), and by \cite{Maja2010} from OGLE RR Lyrae variables (8.1 kpc). Also using RR Lyrae stars (statistical parallaxes), \cite{Dam2009} found a shorter distance of 7.58 kpc. On the other hand, \citet{Ghez2008} and \citet{Gill2009} derived a value of, respectively, 8.4 and 8.33 kpc in monitoring stellar orbits around the massive black hole in the GC. These results are similar to those of \cite{Matsu2009} from Mira variables (8.24 kpc). From a statistical analysis of the estimates published during the last 20 years, \cite{Mal2012} reports a final value of 7.98 kpc. Possible changes to the distance to the GC do not affect the conclusions of this investigation.\\

\begin{table*}[!ht]
\centering
\begin{tabular}{r@{ }lccr@{ }lccr@{ }lcc}
\hline\hline
\multicolumn{2}{c}{Star} & R$_{G}$ & $\sigma$-${\rm R}_{G}$ & \multicolumn{2}{c}{Star} & R$_{G}$ & $\sigma$-${\rm R}_{G}$ & \multicolumn{2}{c}{Star} & R$_{G}$ & $\sigma$-${\rm R}_{G}$ \\
\multicolumn{2}{c}{    } &  kpc    &  kpc                   & \multicolumn{2}{c}{    } &  kpc    &  kpc                   & \multicolumn{2}{c}{    }  &  kpc    &  kpc             \\
\hline                                                       
  AA & Gem & 11.454 & 0.459 & DR & Vel &  8.054 & 0.439 &   TX & Mon & 11.790 & 0.452 \\
  AD & Gem & 10.662 & 0.455 & EK & Mon &  9.960 & 0.453 &   TY & Mon & 11.180 & 0.451 \\
  AD & Pup & 10.588 & 0.434 & EZ & Vel & 12.119 & 0.358 &   TZ & Mon & 11.183 & 0.451 \\
  AH & Vel &  8.074 & 0.450 & HW & Pup & 13.554 & 0.436 &   UX & Car &  7.698 & 0.444 \\
  AO & Aur & 11.835 & 0.461 &  l & Car &  7.845 & 0.451 &   UY & Mon & 10.539 & 0.459 \\
  AO & CMa & 10.430 & 0.433 & MY & Pup &  8.096 & 0.450 &   UZ & Sct &  5.309 & 0.448 \\
  AP & Pup &  8.234 & 0.449 & RS & Ori &  9.470 & 0.453 & V340 & Ara &  4.657 & 0.427 \\
  AQ & Car &  7.658 & 0.425 & RS & Pup &  8.585 & 0.444 & V397 & Car &  7.679 & 0.447\\
  AQ & Pup &  9.472 & 0.436 & RY & CMa &  8.796 & 0.450 & V495 & Mon & 12.098 & 0.453 \\
  AT & Pup &  8.484 & 0.445 & RY & Vel &  7.774 & 0.432 & V508 & Mon & 10.714 & 0.452 \\
  AV & Sgr &  5.980 & 0.454 & RZ & CMa &  9.162 & 0.448 & V510 & Mon & 12.550 & 0.456 \\	 
  AV & Tau & 10.809 & 0.457 & RZ & Gem &  9.973 & 0.454 &    V & Car &  7.915 & 0.447 \\
  AX & Aur & 11.955 & 0.461 & RZ & Vel &  8.249 & 0.445 &    V & Vel &  7.888 & 0.448 \\
  AX & Vel &  8.120 & 0.448 & ST & Tau &  8.897 & 0.452 &   VX & Pup &  8.718 & 0.449 \\
  BE & Mon &  9.609 & 0.452 & ST & Vel &  8.158 & 0.442 &   VY & Car &  7.627 & 0.441 \\
  BG & Vel &  8.000 & 0.446 & SV & Mon & 10.070 & 0.453 &   VY & Sgr &  5.862 & 0.453 \\
  BK & Aur & 10.207 & 0.453 & SW & Vel &  8.457 & 0.433 &   VZ & Pup & 10.867 & 0.425 \\
  BN & Pup &  9.930 & 0.428 & SX & Vel &  8.334 & 0.439 &   WW & Mon & 13.176 & 0.463 \\
  BV & Mon & 10.398 & 0.452 & SY & Aur & 10.271 & 0.454 &   WX & Pup &  9.351 & 0.441 \\
  CS & Ori & 11.701 & 0.458 &  T & Vel &  8.084 & 0.448 &   XX & Mon & 11.854 & 0.451 \\
  CV & Mon &  9.362 & 0.452 & TW & CMa &  9.788 & 0.445 &    Y & Aur &  9.692 & 0.453 \\
     &     &        &       & TW & Mon & 13.059 & 0.457 &   YZ & Aur & 11.668 & 0.459 \\
\hline
\end{tabular}
\caption{Galactocentric distances and uncertainties for the Cepheids in our sample.}
\label{dist}
\end{table*}
    
\subsection{Comparison with V-band-based distances}
    
Because of the recent advent of NIR detectors, the distances of Galactic Cepheids have been traditionally derived from V-band P-L relations (PL$_{V}$). Still, the amount of NIR data available for Galactic Cepheids is less than in the visible part of the spectrum, although the situation is continuously improving. Already in our previous papers \citep{Lem2007,Lem2008,Pedi2009,Pedi2010}, we preferred to use NIR P-L relations: it is now widely accepted that the slopes and zero-points of the P-L relations are less dependent on the metallicity in the NIR than in the visible part of the spectrum. However, it is worth mentioning that no consensus has been reached yet on the amplitude of this so-called metallicity effect, neither from a theoretical nor from an empirical point of view \cite[see, for instance, Fig.~1  in][]{Rom2008}. Moreover, the intrinsic dispersion of the P-L relations is significantly lower in the NIR than in the V-band, leading to lower systematic errors in the NIR. 
    
Finally, another drawback of the P-L$_{V}$-relations has been recently addressed. It is related to the metallicity dependence of the reddening correction: for a long time the main source of reddening values E(B-V) has been the \cite{Fer1995} database, where individual reddening values are computed from a period-color relation not taking into account the metallicity effect, although it is very likely that it has to be considered \citep{LS1994,Bono1999}. The first to adopt a revised version of the Fernie system were \cite{Tamm2003}, and the works of \cite{Laney2007} and \cite{Fou2007} enabled us to correct any reddening value for the metallicity effect. Again, distances derived from NIR P-L relations are less affected by the possible remaining uncertainties in the determination of E(B-V) because the total amount of reddening to be corrected is smaller in this spectral domain.\\
    
   \begin{figure}[!hb]
   \centering
   \includegraphics[angle=0,width=0.98\columnwidth]{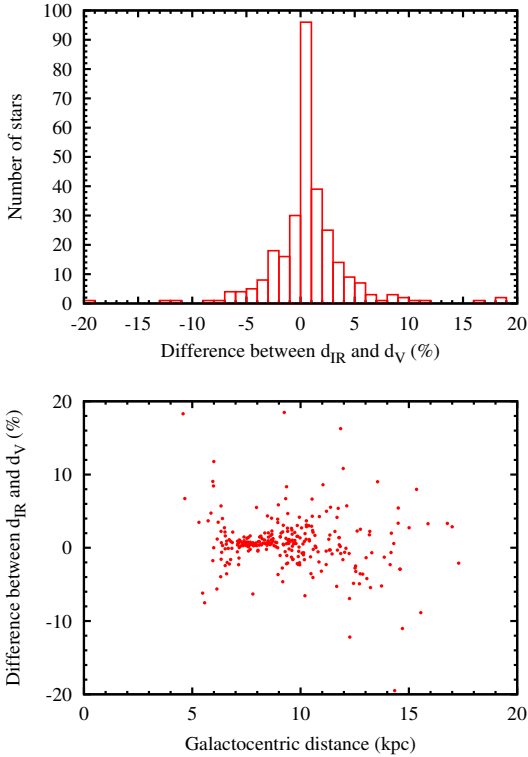}
      \caption{(a) Difference between distances derived from P-W$_{NIR}$ and P-L$_{V}$ relations as a function of the Galactocentric distance of the Cepheid. (b) Comparison between distances derived from P-W$_{NIR}$ and P-L$_{V}$ relations.}
         \label{distcomp}
   \end{figure}

\par In this paper, we adopt distances derived from NIR P-W relations \citep{Inno2013}. This way, we benefit from the advantages of the P-W relations compared to their corresponding P-L relations (see Sect. \ref{Dist}.1) together with those of working in the NIR. As the relations have been derived for Magellanic Clouds Cepheids, we assume that the reddening law is the same in different systems. Fig.~\ref{distcomp}a shows the overall good agreement between distances derived either from the P-L$_{V}$ taken from \citet{Luck2011a}, \citet{Luck2011b}, or from the P-W$_{NIR}$ relations, because the difference is in general less than 5\%. Moreover, the difference is less than 2\% for a significant part of our sample, especially close to the Solar neighborhood, where one can note in Fig.~\ref{distcomp}b the narrow distribution. On the other hand, the discrepancies are increasing when going further from the Sun, both toward the inner and toward the outer disk. Such a result is all but a surprise: the Cepheids located close to the Sun are not expected to be highly reddened, and therefore the discrepancies between P-L$_{V}$ and P-W$_{NIR}$ are expected to be minimal. When moving away from the Sun, one expects not only an increase of the reddening but also (from our previous knowledge of the Galactic metallicity gradient) an increase of the Cepheids' metallicity toward the inner disk and a decrease toward the outer disk. Both these characteristics will significantly affect the accuracy of the distances derived from P-L$_{V}$ relations. The values quoted above are in good agreement with uncertainties at a level of 13\% reported by \cite{Luck2011b}. Here we would like to stress that {\it a discrepancy as small as 5\% at 10 kpc translates into a significant discrepancy of 0.5 kpc}. As the current sample of Cepheids with known abundances is largely located at $\pm$1--2 kpc from the Sun, such an uncertainty on the distances does not yet strongly affect the determination of the gradients (differences $<$0.005 dex/kpc). However, this effect will certainly increase, together with an improved sampling (in terms of chemical composition) of the inner/outer disk or towards other Galactic quadrants, and affect the current conclusions.\\

\section{Results: abundances}
\label{Abund}

\par In this section we report the abundances obtained for the individual Cepheids and compare when possible with previous studies. For stars in common, we computed the mean value and standard deviation of the difference between our results and those of \cite{Luck2011b} and \cite{Luck2011a} for S and Zr \citep[we have only three stars in common with][]{Yong2006}. For a proper comparison, the results of \citet{Yong2006}, \citet{Luck2011a}, and \citet{Luck2011b} have been scaled when needed to the solar chemical abundances of \citet{Gre1996}. Values are given in Table~\ref{diffstand}. As far as [Fe/H] is concerned, we found an excellent agreement (differences $\leq$ 0.1 dex in most cases) between our studies \citep{Lem2007,Lem2008} and the results of \cite{Luck2006}.\\

\subsection{Light elements}

\par {\it Sodium:} Our sodium abundances cover a wide range of values, and, as expected for Galactic supergiants, the Cepheids in our sample are in most cases Na-overabundant. This overabundance is due to the synthesis of sodium via the Ne-Na cycle in the convective cores of the intermediate-mass main sequence stars, which are the progenitors of Cepheids. After the first dredge-up, mixing brings the Na-enriched material into the radiative envelope of the stars \citep[e.g.,][and references therein]{Sass1986,Deni1994,Take2013}. Our results are in good agreement with \cite{Luck2011b}.\\ 

\noindent {\it Aluminium:} Cepheids have aluminium abundances in the [--0.3~dex~--~+0.3 dex] range. Once again, our results are in good agreement with \cite{Luck2011b}. Although there are 6 Al lines in our line list, they are usually weak in Cepheids and our results typically rely on 2--3 lines.\\

\subsection{$\alpha$ elements}

\par {\it Magnesium:} Although magnesium abundances have been found mostly sub-solar in all of the previous Cepheid studies, \cite{Luck2011b} report higher [Mg/H] values. The impact of the revision varies from star to star but as a whole can be considered as a shift upward of $\approx$0.2 dex. The method is exactly the same as in previous studies (including the value of [Mg/H]$_{\odot}$), and the possible systematics, which are examined thoroughly by \cite{Luck2011b}, are extremely small. Moreover, \cite{Luck2011b} have several dozens of stars in common with previous studies, so the shift of the Mg abundances cannot be related to azimuthal inhomogeneities in the Milky Way disk. Therefore it is hard to imagine another cause to explain the shift than the new line selection used by \cite{Luck2011b}.
\par In our study we retained two Mg lines at 5711.09~\AA{} and 8736.02~\AA,{} respectively. However, the latter could not be observed in all of our spectra as the efficiency of the optical spectrographs (and in turn the S/N) decreases rapidly when approaching the NIR. We did not retain the Mg triplet around 6318~\AA. These lines are close to an auto-ionization Ca line, which makes it harder to properly determine the continuum placement. This determination is crucial as these lines are generally weak in Cepheids. Because the 8736~\AA{} line is also a multiplet, it may also be better not to include it in our line list. On the other hand, Mg abundances derived from this line have always been consistent with those from the 5711~\AA{} line when both could be measured in the Cepheids of our sample.
\par Recently, \cite{Merle2011} studied the NLTE effects on some Mg and Ca lines in late-type giants/supergiants. For the 8736~\AA{} line, they found that the NLTE correction does not exceed --0.1 dex in the metallicity and gravity range of Galactic Cepheids. However, the temperature of their models does not exceed 5250 K, while Galactic Cepheids have temperatures ranging from 4800 K to 6200 K.  As far as the 5711~\AA{} line is concerned and for the same temperature range, NLTE effects are negligible for Galactic Cepheids and are noticeable only below [Fe/H]~=~--~1.5 dex.
\par Our results support the findings of \cite{Luck2011b}: Cepheids in our sample have abundances ranging mostly from --0.3 dex to +0.3 dex, except toward the inner disk, where they reach higher values. They also fall on the same locus in the [Mg/H] vs R$_{G}$ plane as the Cepheids in \cite{Luck2011b} and show a similar dispersion at a given Galactic radius (Fig.~\ref{gradient_alpha}). The same applies to the Cepheids of \cite{Yong2006}, for which we could compute NIR-based distances.\\

\noindent {\it Silicium:} Our silicium abundances are in excellent agreement with those of \citet{Luck2011a}, \citet{Luck2011b}, and previous studies. This is not surprising as a larger number of good Si lines (typically 10--12) can be measured in Cepheids.\\

\noindent {\it Sulfur:} Our results show a greater scatter than for the other $\alpha$ elements, similar to what was found for sulfur in other studies \citep{Luck2011a}. This is certainly due to the fact that our [S/H] abundances are based on typically 1-2 relatively weak lines. In addition to this scatter, it seems that our sulfur abundances are shifted upward by 0.1--0.2 dex, compared to those of \cite{Luck2011a}. As for Mg, this could be due to the line selection because we include in our line list the lines at 6052.67~\AA{} and 6538.60~\AA,{} which are not in the list of \cite{Luck2011a}. In contrast, however, they retained the lines at 7679.60~\AA{} and 7686.03~\AA,{} which we did not measure. It is worth mentioning that we found no correlation between the [S/H] scatter and the longitude.
\par It would be very important to reduce the scatter and to ascertain the sulfur abundances of Cepheids. Unlike C-N-O, sulfur is not processed in the intermediate and massive stars and thus traces the sulfur abundance in the interstellar medium at the time the Cepheid was born. This would enable a direct gradient comparison between Cepheids and the other tracers of the young population.\\ 

\noindent {\it Calcium:} Although there are also many Ca lines in the Cepheid spectra, they are quite often strong lines and depart from a Gaussian profile, which makes them less suitable for abundance determination. At the end, we have typically 5-7 good Ca lines that have been properly measured. The dispersion is similar to the one obtained for Si. The agreement is also very good with previous studies. Among the lines in our list that have been studied by \cite{Merle2011}, only the one at 6166.44~\AA{} suffers from a noticeable NLTE correction approaching 0.1 dex at T$_{eff}$=5250 K.\\

\subsection{Heavy elements}

{\it Europium:} Our europium abundances are in good agreement with those from \cite{Luck2011b}. With the exception of a few stars with rather large discrepancies, the differences are in general of the order of 0.15 dex or less. In both studies, [Eu/H] is derived from only two lines, of which only one (at 6645.13~\AA) is in common.\\

\noindent {\it Other heavy elements (Y, Zr, La, Ce, Nd):} We discuss the other heavy elements together as our comments will be very similar for all of them. When compared to \cite{Luck2011b}, our measurements agree well for roughly half of our sample, which are mostly metal-rich Cepheids that are $\pm$2 kpc from the Sun. The other half are in general more metal-poor Cepheids located beyond $\approx$ 10 kpc, and our heavy elements abundances are very often lower than those of \cite{Luck2011b}. The discrepancies are largest in the case of Y, La, and Nd, while Ce abundances agree very well. Because Zr has not been studied by \cite{Luck2011b}, we compare with \cite{Luck2011a}, and our Zr abundances are in much better agreement than those of the other heavy elements.   
\par The origin of this ``bimodal'' behavior is not clear. We have already mentioned that our approaches are very similar. 
Different causes can be invoked to understand this behavior: 
\begin{itemize}
	\item It could arise from systematics in the determination of the atmospheric parameters of the Cepheids. However, we found an excellent agreement for [Fe/H] and most of the other elements we analyzed. Moreover, this could not explain why our heavy element abundances agree in general relatively well for the more metal-rich stars in our sample and not for the most metal-poor ones. We thus discard this possibility.
	\item It seems that we have to look in the direction of the absorption lines of the heavy elements. However, we can also exclude that the measurement errors (like a wrong fit of the lines or errors in the continuum placement) are larger for the lines of the heavy elements than in the case of other species. Moreover, it is easier to properly locate the continuum in the Cepheids that have sub-solar metallicities than in those that are more metal-rich than the Sun because the line blending decreases with metallicity. On the other hand, the more metal-poor stars in our sample are found toward the outer disk (a consequence of the abundance gradients) and therefore have lower S/N due to their greater distances.
	\item We note that our analysis of heavy elements is based on a slightly larger number of lines than the other studies: We typically measure 4-6 Y lines, 1 Zr line, 4-6 La lines, 2-3 Ce lines, and 2-4 Nd lines, while the line list of \cite{Kov1999} lists only 2 lines for each heavy element. For the lines that are in common, we checked that we have similar atomic parameters, except for the Y line at 6795.41~\AA. For this Y line we used a {\it log~gf} value of --1.03, while \cite{Kov1999} used a value of --1.20. \cite{Yong2006} measured one Eu line and four La lines, among which three are in common with our own line list and for which they retained the same atomic parameters. We also note that the differences are greater for the elements where we used a larger number of lines than \cite{Luck2011b} (Y, La, Nd), while the agreement is quite good when we used a similar number of lines (Zr, Ce, Eu).  
	\item As the atomic parameters for the lines in common are identical or very similar in most cases, it seems unlikely that they account for much of the difference. It is probable that at least some of the lines of these elements (which are, moreover, found in their ionized state) are altered by NLTE effects. Selecting different lines (that are not affected in the same way by NLTE effects) in different line lists could then lead to the discrepant results we observe. None of the studies compared here (\cite{Yong2006,Luck2011a,Luck2011b}; this study) computed NLTE effects for the heavy elements or took into account hyperfine structure corrections. Repeating with other elements the NLTE analysis of barium in Milky Way Cepheids made by \cite{And2013} would certainly help to clarify the situation.
\end{itemize}

\begin{table}[!ht]
\centering
\begin{tabular}{rcc}
\hline\hline
Element & Mean difference &  $\sigma$ \\
        &      dex        &   dex  \\
\hline
{[}Na/H] & 0.12 & 0.11 \\
{[}Mg/H] & 0.19 & 0.15 \\
{[}Al/H] & 0.13 & 0.09 \\
{[}Si/H] & 0.11 & 0.07 \\
{[}S/H] & 0.18 & 0.16 \\
{[}Ca/H] & 0.14 & 0.11 \\
{[}Y/H] & 0.16 & 0.11 \\
{[}Zr/H] & 0.17 & 0.13 \\
{[}La/H] & 0.20 & 0.14 \\
{[}Ce/H] & 0.14 & 0.11 \\
{[}Nd/H] & 0.20 & 0.12 \\
{[}Eu/H] & 0.18 & 0.16 \\
\hline
\end{tabular}
\caption{Mean value and standard deviation of the abundance difference between this study and \cite{Luck2011b}, except for S and Zr, where the 
comparison is made with \cite{Luck2011a}.}
\label{diffstand}
\end{table}

\section{Results: gradients}
\label{Res}

   \begin{figure*}[!ht]
   \centering
   \includegraphics[scale=0.85]{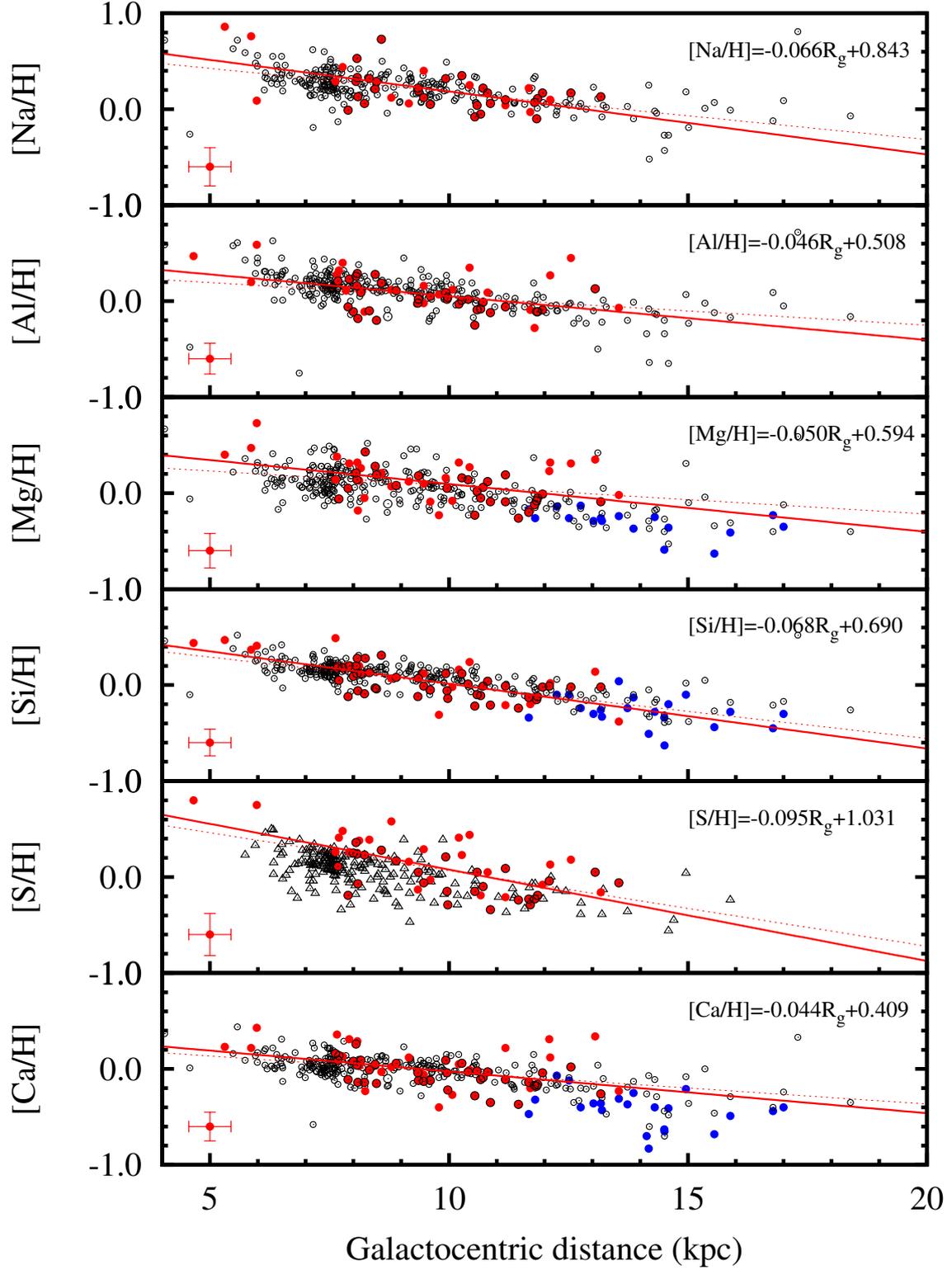}   
   \caption{Galactic abundance gradients from Cepheids for light and $\alpha$ elements. The full red dots are our data. We also show data from \citet{Luck2011b} (black open circles) and if not available, from \citet{Luck2011a} (black open triangles). When stars are in common and the abundances fall within 0.2 dex, we overplotted black open circles (triangles) to the red dots. \cite{Yong2006} also provided data for several elements in the outer disk (full blue dots). All the distances are derived with a PW$_{NIR}$ relation. A representative error bar is shown, including both the uncertainty on the atmospheric parameters and the rms uncertainty on the mean abundance [X/H]. We also indicate a linear fit to our data on the [4-15] kpc range (thick red line) and (for some elements) on the [7-15] kpc range (dashed red line).}
         \label{gradient_alpha}
   \end{figure*}
%

   \begin{figure*}[!ht]
   \centering      
   \includegraphics[scale=0.85]{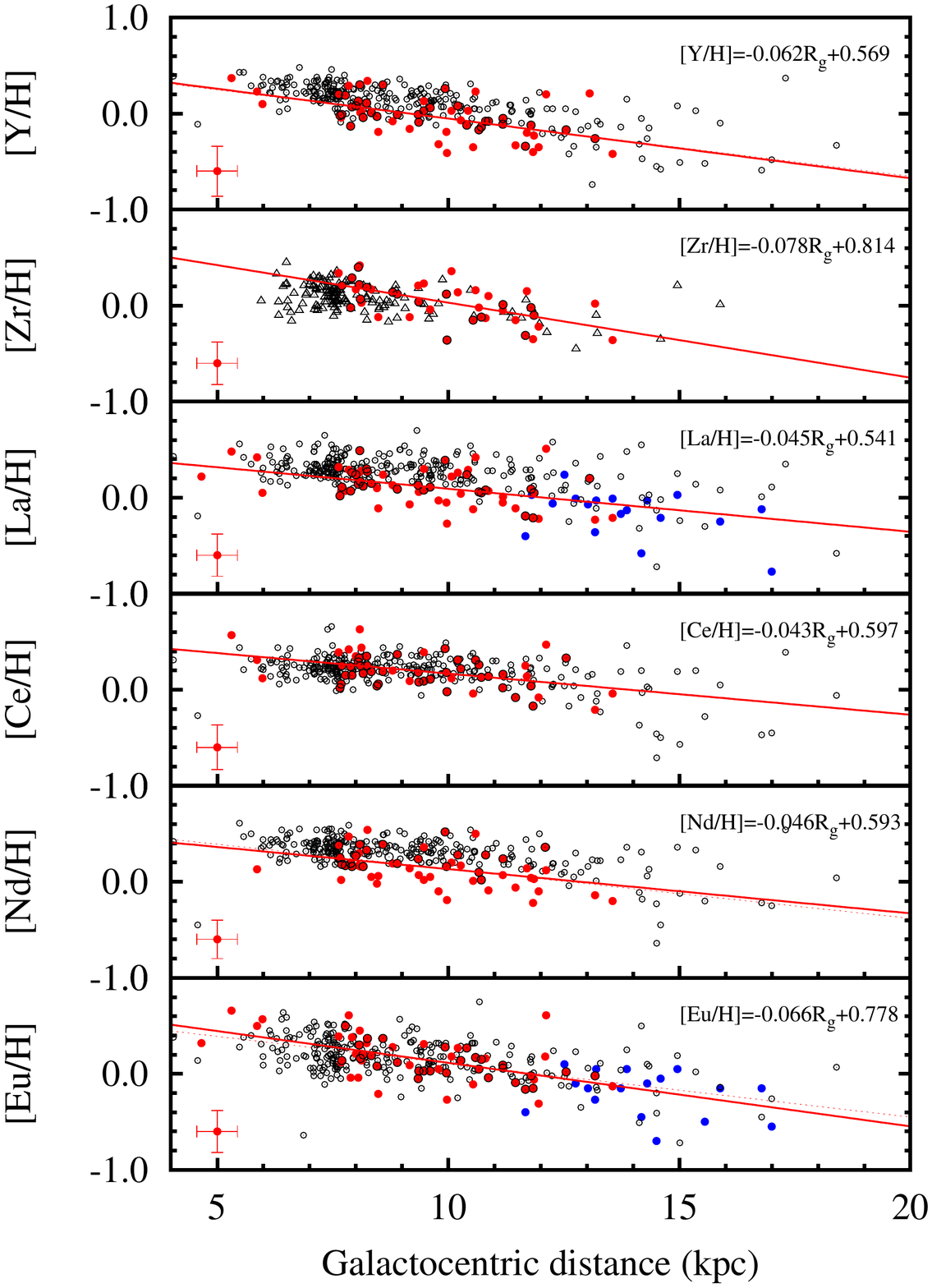}
   \caption{Same as Fig.~\ref{gradient_alpha}, but for heavy elements.}
         \label{gradient_heavy}
   \end{figure*}

\par When examining the abundance trends with Galactocentric distance, we first considered simple, linear gradients. Fig.~\ref{gradient_alpha} displays linear gradients for several light (Na, Al) and $\alpha$ elements, while Fig.~\ref{gradient_heavy} shows the gradients for the heavy elements. Details about the linear regression are given in Table~\ref{gradients}.\\

\begin{table*}[!ht]
\centering
\begin{tabular}{rccccccc}
\hline\hline
Element  & range  &    slope          &   zero-point     &  rms  &  N & slope (previous studies) &  rms   \\
         &  kpc   &  dex/kpc          &      dex         &       &    &      dex/kpc             &        \\
\hline
{[}Na/H] & [4-15] & --0.066 $\pm$0.015 & 0.843 $\pm$0.146 & 0.177 & 42 & --0.047 $\pm$0.003 & 0.131 \\
         & [7-15] & --0.049 $\pm$0.016 & 0.673 $\pm$0.161 & 0.154 & 39 &                    &       \\
{[}Mg/H] & [4-15] & --0.050 $\pm$0.013 & 0.594 $\pm$0.133 & 0.185 & 54 & --0.048 $\pm$0.004 & 0.160 \\
         & [7-15] & --0.030 $\pm$0.015 & 0.381 $\pm$0.150 & 0.174 & 51 &                    &       \\
{[}Al/H] & [4-15] & --0.046 $\pm$0.013 & 0.508 $\pm$0.122 & 0.170 & 55 & --0.049 $\pm$0.003 & 0.130 \\
         & [7-15] & --0.030 $\pm$0.014 & 0.343 $\pm$0.142 & 0.164 & 52 &                    &       \\
{[}Si/H] & [4-15] & --0.068 $\pm$0.009 & 0.690 $\pm$0.089 & 0.139 & 62 & --0.048 $\pm$0.002 & 0.088 \\
         & [7-15] & --0.057 $\pm$0.011 & 0.575 $\pm$0.110 & 0.140 & 58 &                    &       \\
{[}S/H]  & [4-15] & --0.095 $\pm$0.015 & 1.031 $\pm$0.152 & 0.206 & 52 & --0.076 $\pm$0.005 & 0.159 \\
         & [7-15] & --0.079 $\pm$0.017 & 0.857 $\pm$0.173 & 0.202 & 50 &                    &       \\
{[}Ca/H] & [4-15] & --0.044 $\pm$0.012 & 0.409 $\pm$0.119 & 0.173 & 61 & --0.041 $\pm$0.003 & 0.114 \\
         & [7-15] & --0.033 $\pm$0.014 & 0.302 $\pm$0.138 & 0.173 & 58 &                    &       \\
{[}Y/H]  & [4-15] & --0.062 $\pm$0.012 & 0.569 $\pm$0.114 & 0.166 & 61 & --0.061 $\pm$0.003 & 0.116 \\
{[}Zr/H] & [4-15] & --0.078 $\pm$0.016 & 0.814 $\pm$0.158 & 0.162 & 41 & --0.047 $\pm$0.005 & 0.127 \\
{[}La/H] & [4-15] & --0.045 $\pm$0.012 & 0.541 $\pm$0.111 & 0.168 & 60 & --0.031 $\pm$0.004 & 0.144 \\
{[}Ce/H] & [4-15] & --0.043 $\pm$0.012 & 0.597 $\pm$0.113 & 0.158 & 55 & --0.034 $\pm$0.003 & 0.133 \\
{[}Nd/H] & [4-15] & --0.046 $\pm$0.013 & 0.593 $\pm$0.131 & 0.165 & 55 & --0.037 $\pm$0.003 & 0.114 \\
{[}Eu/H] & [4-15] & --0.066 $\pm$0.013 & 0.778 $\pm$0.122 & 0.184 & 59 & --0.042 $\pm$0.005 & 0.171 \\
         & [7-15] & --0.056 $\pm$0.015 & 0.671 $\pm$0.153 & 0.184 & 55 &                    &       \\
\hline
\end{tabular}
\caption{Slopes and zero-points of the abundance gradients for various elements. We also indicate the root mean square (rms) of the residuals and the number of data points for the fit. For some elements, we also computed the gradient over a different distance range. The gradient values and rms obtained by \cite{Luck2011b} are tabulated for comparison (\cite{Luck2011a} for S and Zr).}
\label{gradients}
\end{table*}

\par Because Na and possibly Al are modified prior to the Cepheid evolutionary stage, their gradients should only be considered as indicative. For Na, the high value obtained for the slope of the gradient (--0.066 dex/kpc) is largely due to the influence of a couple of Na--rich stars in the inner disk. If in contrast we compute the gradient only in the range [7-15]~kpc, the value we obtain (--0.049 dex/kpc) is very similar to the results of \cite{Luck2011a} and \cite{Luck2011b}, which are --0.044 dex/kpc and --0.047 dex/kpc, respectively. Our results are also very similar for the Al gradient: we found a slope of --0.046 dex/kpc over the whole Galactic disk, while \cite{Luck2011a}, \cite{Luck2011b} gave slopes of --0.053 and --0.49 dex/kpc, respectively. This value is also similar to the slope of $-0.048$ dex/kpc found by \cite{Daf2004b} from OB stars.\\

\par For the $\alpha$ elements, Fig.~\ref{gradient_alpha} shows that our abundance trends with Galactocentric distance agree very well with the studies of \cite{Luck2011a}, \cite{Luck2011b}, and \cite{Yong2006}. Indeed, the locus of the stars in the [M/H] vs. R$_{G}$ plane and the dispersion around the slope match very well. The actual value of the slope (tabulated in Table~\ref{gradients}) might be affected in some cases by the 3--4 $\alpha$ rich stars in the inner disk and/or by 2--4 stars at 12--14 kpc with higher $\alpha$-abundances than the other Cepheids at these Galactocentric distances. In particular for Mg, the best fit seems to be artificially shifted by $\approx$+0.2 dex from the rest of the sample under the combined influence of these stars. From the figure it seems, however, that the slope is correct, and indeed our computed slope (--0.050 dex/kpc) is in excellent agreement with the results of \cite{Luck2011a}, and \cite{Luck2011b}, who both found a slope of --0.048 dex/kpc, and with those of \cite{Daf2004b}, who found a slope of --0.052 dex/kpc. We already mentioned that our [S/H] abundances are higher by +0.1 -- +0.2~dex than those of \cite{Luck2011a}, which is confirmed by Fig.~\ref{gradient_alpha}. We find, however, a similar slope (--0.079 dex/kpc vs. --0.076 dex/kpc) if we ignore the two inner disk Cepheids, while \cite{Daf2004b} report a much shallower slope (--0.040 dex/kpc). The dispersion is much lower for Si and, to a slightly lesser extent, for Ca because their abundances are determined with a larger number of lines. Given the size of the respective error bars, this dispersion is certainly closer to reflecting a real star-to-star scatter, while it is probably dominated by the uncertainties of the abundance determinations for the other elements. Our Ca slope (--0.044 dex/kpc) is very close to those determined by \cite{Luck2011a} and \cite{Luck2011b}, which are --0.040 dex/kpc and --0.041 dex/kpc,  respectively. It is surprising to note that the inclusion (or not) of the four slightly more Si-rich stars in the inner disk noticeably changes the value of the slope, from --0.068 dex/kpc to --0.057 dex/kpc. Both these values are larger than those reported by \cite{Luck2011a} and \cite{Luck2011b} (--0.049 dex/kpc and --0.048 dex/kpc, respectively), and the difference is even larger with \cite{Daf2004b}, who found a slope of only --0.040~dex/kpc. The lower value for the slope has our preference as it better fits the data of \cite{Luck2011a} and \cite{Luck2011b}, while the abundances of the inner disk Cepheids do not deviate much from this slope. On the other hand, the higher values better fits the data of \cite{Yong2006}.\\
\indent The slope values are also in good (Mg, Si) or very good (Ca) agreement with those derived by \cite{Ces2007} in their chemical evolution models, which are --0.039, --0.045, --0.047~dex/kpc for Mg, Si, and Ca, respectively. However, they report for sulfur a similar slope (--0.047 dex/kpc) that is much shallower than the ones found in this study or the study by \cite{Luck2011a}. The steeper slope found for sulfur should then be regarded with some caution, and the determination of the sulfur abundances in Cepheids clearly deserves further investigation.\\
\indent \cite{Ces2007} also report that the  $\alpha$ element abundances predicted by their model around the solar circle are slightly higher than the ones reported at that time from Cepheid observation, and that the difference is much greater for Mg. In this respect, the upward revision of the Mg abundances by \cite{Luck2011b}, supported here, improves the agreement between the model and the observations. From a different point of view, \citet{Pio2012} constructed a model with radial flows of dust and gas but no radial migration of stars. The model reproduces well the observations of gradients from different tracers, in particular the Cepheids, for a good number of proton-capture and $\alpha$ elements.\\

\par The existence of radial abundance gradients for the heavy elements has been more controversial than for the other elements. Their spectral lines are not very numerous, either rather weak or too strong (especially in the case of Ba), and quite often located in crowded parts of the spectra, which involves blending. These difficulties lead to an increased observational scatter and hence to inconclusive results in terms of gradients. It is only in the recent studies of \citet{Luck2011a} and \citet{Luck2011b} that the existence of gradients has been recognized for all of the heavy elements measured. On the other hand, \cite{And2013} found no gradient at all in a recent NLTE study of barium in Cepheids over the whole Galactic disk.\\
\indent Our results confirm the existence of radial abundance gradients (within the context of an LTE analysis) for all the heavy elements we studied (Y, Zr, La, Ce, Nd, Eu). For Ce, the locus of our points and the value of the slope are in good agreement with \citet{Luck2011b}, while for Zr the slope is overestimated (despite a similar locus) when compared to \citet{Luck2011a}, probably due to the absence of Zr measurements below 7.5 kpc in our Cepheid sample. The europium trend with Galactocentric distance also seems to be in very good agreement when comparing our results to the observations of \cite{Luck2011b} and \cite{Yong2006} in Fig.~\ref{gradient_heavy}. It is then surprising that the slopes are not in good agreement (--0.066 vs. --0.042 dex/kpc). Again, this is probably linked to the large number of stars with a solar-like abundance around 7 kpc in the sample of \cite{Luck2011b}, which we do not have in our sample, together with the influence of the Eu-rich inner disk Cepheids. Indeed, when the latter are not taken into account, the slope goes down to --0.056 dex/kpc. As far as the other elements are concerned, the slopes are in good agreement (Y, Nd) with previous results, despite the differences in abundance results discussed in Sect.~\ref{Abund}. The discrepancies are largest in the case of La, and for this element our gradients match much better to the outer disk Cepheids of \citet{Yong2006} than to the large sample of \citet{Luck2011b}. We believe that this effect is related to the La absorption lines that are included or not in the line lists used by the different authors. The slope we obtain (--0.045 dex/kpc) is then much steeper than the --0.031~dex/kpc reported by \citet{Luck2011a}. \cite{Ces2007} noted that the La abundances in the solar neighborhood of \cite{Luck2006} were higher than those predicted by their model. As a result, the slope predicted by the model (-0.032 dex/kpc in the [4-14] kpc range) was steeper than the one seen in the data available at that time. It is, however, close to the current value proposed by \citet{Luck2011b}. Considering only the objects located within 13 kpc from the Galactic center, \citet{Yong2012} compared the slope of the gradients for the Cepheids and for the young ($<$1 Gyr) open clusters in their compilation (see their Fig. 32). The slopes are in very close agreement for all of the elements except La, where the difference reaches $\approx$0.05 dex (the open clusters gradient has a larger value than the Cepheid gradient). It should be noted that the uncertainties on the La gradient from open clusters are also much greater than for any other element. Clearly La deserves further investigation, both from a Cepheid and from an open cluster point of view. While Y and Nd show a similar behavior to La, the differences are less pronounced and the slopes of our gradients are in good agreement with \citet{Luck2011b}.\\

\par A steeper slope for the iron gradient in Cepheids toward the inner disk was first reported by \citet{And2002b} and supported by \citet{Pedi2009,Pedi2010} and more recently on a larger sample by \citet{Geno2013a}. On the basis of the current (small) sample of Cepheids towards the inner disk, it is impossible to say if such a distinctive feature also holds for the other elements: the shallower slopes found when the inner disk Cepheids are discarded seem to speak in favor of this scenario; on the other hand this could simply be caused by the severe undersampling of inner disk Cepheids. We plan to focus on this region in more detail in a forthcoming paper \citep{Geno2013b}.

\par In contrast with previous studies, we found no clear evidence of a flattening of the gradient in the outer disk, as was already pointed out by \citet{Luck2011a} and \cite{Luck2011b}. This is in contrast to the young population in distant galaxies and to the open clusters in the Milky Way. \citet{Jaco2011b}, \citet{Yong2012}, and most of the previous open clusters studies report a flattening of the gradient in the outer disk. When splitting the sample in three age bins, \citet{Jaco2011b} report a possible flattening of the gradient, but also a possible change in the location of the transition zone with time. The outer disk flattening seems to occur around 10 kpc for the oldest clusters, while [Fe/H] in the intermediate-age clusters decreases linearly and reaches the same ceiling value only around 14 kpc (where the distribution of the intermediate-age clusters stops).
It is impossible to draw any firm conclusion yet due the unequal
sampling of each bin (each age bin contains a different number of
clusters and the sampling in terms of Galactocentric distance is also
different). Because the recent studies could not detect a flattening
of the Cepheid gradient in the outer disk, the Cepheids would then
easily fit in that context. A comprehensive comparison between
Cepheids and open clusters has been made by \citet{Yong2012}; we will
not repeat the exercise here. The simulations of \cite{Min2012} also
show no flat gradient in the outer disk for the young stars. They
found a strong flattening of the gradient for the old stars, which was also found by \citet{Wang2013}, but the young populations are much less affected. As a result, the initial and the present-day gradient are very similar out to $\approx$12 kpc, where some flattening occurs. Indeed, the differences between the initial/final slopes are very small and anyway smaller than the uncertainties on the observed gradient.\\

\par Large surveys combining photometry and spectroscopy enable us to derive the radial velocities and the atmospheric parameters of numerous stars. The size of the samples and their high internal consistency compensates for the somewhat less accurate abundances and distances compared to the tracers commonly used to study the abundance gradients in the Milky Way. From the Geneva-Copenhagen Survey, \citet{Nord2004} derived a value of --0.076$\pm$0.014 dex/kpc for stars younger than 1.5 Gyr in the solar neighborhood ($\approx$ [7--9] kpc). From the third data release (DR3) of the RAVE survey \citep{Stein2006}, \citet{Cosku2012} derived for young, F-type stars a radial metallicity [M/H] gradient of --0.051$\pm$0.005~dex/kpc similar to the Cepheid gradient. Again from the RAVE DR3 data but this time for red clump stars in the thin disk, \citet{Bilir2012} report a slightly shallower gradient of [M/H] = --0.041$\pm$0.005 dex/kpc. From MSTO stars (which are older than Cepheids) in the SEGUE data \citep{Yanni2009}, \cite{Cheng2012a} found a gradient of --0.066 dex/kpc for stars at 0.15 kpc $<$~$|$z$|$~$<$ 0.25 kpc, with a gradient flattening toward higher altitudes. This value is close to the one obtained for Cepheids. However, it should be noted that most of the Cepheids are located below $|$z$|$ = 0.15 kpc. Moreover, the SEGUE subsample considered contains several hundreds of stars in the [0.15--0.25] kpc layer, but over a range of Galactocentric distances limited to the [7--10] kpc region. Taking advantage of a large spectroscopic sample collected in preparation for the CoRoT mission \citep{Loei2008}, \citet{Gazza2013} determined the atmospheric parameters of numerous Milky Way disk stars with the help of the MATISSE algorithm \citep{Recio2006}. They report a radial metallicity gradient of --0.097$\pm$0.015 dex/kpc with giant stars and --0.053$\pm$0.015 dex/kpc with dwarfs; the latter in very good agreement with the value obtained from Cepheids. Most of the results quoted above concern the iron gradient and sometimes also the $\alpha$ elements \citep{Cheng2012b,Gazza2013}. Possible comparisons for a larger number of elements await the results of the ongoing large spectroscopic surveys like APOGEE \citep{Maje2007}, the GAIA-ESO survey \citep{Gil2012}, or HERMES \citep{Free2012}.

\section{Conclusions}
\label{Con}

\par We used high-resolution spectra (FEROS, ESPADONS, NARVAL) to determine the abundances of several light (Na, Al), $\alpha$~(Mg, Si, S, Ca), and heavy elements (Y, Zr, La, Ce, Nd, Eu) in a sample of 65 Milky Way Cepheids. Combining these results with accurate distances from p-W relations in the NIR enables us to study the abundance gradients in the Milky Way. Our results are in good agreement with previous studies from Cepheids or other tracers. In particular:
\begin{itemize}
	\item We confirm an upward shift of $\approx$0.2 dex for the Mg abundances first reported by \citet{Luck2011b}.
	\item The locus of our points is for most of the elements in good agreement with previous Cepheid studies \citep{Yong2006,Luck2011a,Luck2011b}. The slopes are sometimes discrepant, mainly due to sampling effects, as a few points sometimes have a strong influence on the determination of the slope. Our sample, originally designed to study either the outer disk or the very inner disk, lacks targets in the solar neighborhood.
	\item We confirm the existence of a gradient for all the heavy elements in the context of an LTE analysis. Moreover, for Y, Nd, and especially La, we find lower abundances for Cepheids in the outer disk than reported in previous studies, leading to steeper slopes for the gradients. We attribute this effect to the differences in the line lists used by different groups. On the other hand, \citet{And2013} report no gradient for an NLTE analysis of Ba. Further investigations are needed to assess the behavior of the heavy elements and quantify, in particular, the role of the asymptotic giant branch stars in the chemical evolution of the Milky Way.
	\item Current data do not support a flattening of the gradients in the outer disk, in agreement with the observations of \citet{Luck2011a}, \cite{Luck2011b}, and the simulations of \citet{Min2012}. This is different from what is observed in open clusters, but remains coherent with a picture where the transition zone between the inner disk and the outer disk would move outward with time \citep{Jaco2011b}. 
\end{itemize}

\begin{acknowledgements}
We thank the anonymous referee for the careful reading of our manuscript and the valuable comments.
This work was partially supported by PRIN-INAF 2011 ``Tracing the formation
and evolution of the Galactic halo with VST'' (PI: M. Marconi) and by
PRIN-MIUR (2010LY5N2T) ``Chemical and dynamical evolution of the Milky Way
and Local Group galaxies'' (PI: F. Matteucci).
G.B. thanks ESO for support as science visitors. 
V.K. is grateful for the support from the Swiss National Science Foundation, project SCOPES No. IZ73Z0-128180.  
B.L. thanks the French Programme National de Physique Stellaire (PNPS) for the financial support of the observing missions.
\end{acknowledgements}

\begin{appendix}
\section{Line list}
\begin{table}[!ht]
\centering
\scriptsize
\begin{tabular}{cccr}
\hline\hline
Wavelength   & Element  & $\chi_{ex}$  & $log~gf$ \\
  $\AA$      &          &     eV       &          \\
\hline
 7835.31 & Al I   & 4.02   &   --0.65  \\
 7836.13 &  Al I  & 4.02   &   --0.49  \\
 4486.91 &  Ce II & 0.29   &   --0.26  \\
 4554.04 &  Ba II & 0.00   &    0.17  \\
 4562.37 &  Ce II & 0.48   &    0.23  \\
 4578.56 &  Ca I  & 2.52   &   --0.70  \\
 4685.27 &  Ca I  & 2.93   &   --0.88  \\
 4721.02 &  Ca II & 7.05   &   --0.78  \\
 4959.12 &  Nd II & 0.06   &   --0.80  \\
 5044.02 &  Ce I  & 1.21   &   --0.07  \\
 5092.79 &  Nd II & 0.38   &   --0.61  \\
 5112.28 &  Zr II & 1.66   &   --0.85  \\
 5114.56 &  La II & 0.23   &   --1.03  \\
 5119.12 &   Y II & 0.99   &   --1.36  \\
 5130.59 &  Nd II & 1.30   &    0.45  \\
 5181.17 &  Nd II & 0.86   &   --0.74  \\
 5289.81 &   Y II & 1.03   &   --1.85  \\
 5290.82 &  La II & 0.00   &   --1.65  \\
 5349.47 &  Ca I  & 2.71   &   --0.31  \\
 5402.77 &   Y II & 1.84   &   --0.63  \\
 5431.52 &  Nd II & 1.12   &   --0.47  \\
 5509.91 &   Y II & 0.99   &   --0.95  \\
 5518.49 &  Ce II & 1.15   &   --0.67  \\
 5581.97 &  Ca I  & 2.52   &   --0.55  \\
 5590.11 &  Ca I  & 2.52   &   --0.57  \\
 5601.28 &  Ca I  & 2.53   &   --0.52  \\
 5665.56 &  Si I  & 4.92   &   --1.94  \\
 5688.21 &  Na I  & 2.10   &   --0.40  \\
 5690.43 &  Si I  & 4.93   &   --1.77  \\
 5711.09 &  Mg I  & 4.35   &   --1.83  \\
 5728.89 &   Y II & 1.84   &   --1.12  \\
 5805.77 &  La II & 0.13   &   --1.56  \\
 5853.69 &  Ba II & 0.60   &   --0.91  \\
 5867.56 &  Ca I  & 2.93   &   --1.57  \\
 5948.54 &  Si I  & 5.08   &   --1.13  \\
 6043.39 &  Ce II & 1.21   &   --0.48  \\
 6125.03 &  Si I  & 5.61   &   --1.46  \\
 6141.71 &  Ba II & 0.70   &   --0.03  \\
 6154.23 &  Na I  & 2.10   &   --1.55  \\
 6155.14 &  Si I  & 5.62   &   --0.75  \\
 6160.75 &  Na I  & 2.10   &   --1.25  \\
 6166.44 &  Ca I  & 2.52   &   --1.14  \\
 6244.48 &  Si I  & 5.62   &   --1.09  \\
 6262.29 &  La II & 0.40   &   --1.22  \\
 6300.30 &   O I  & 0.00   &   --9.71  \\
 6363.77 &   O I  & 0.02   &   --10.2  \\
 6390.48 &  La II & 0.32   &   --1.41  \\
 6414.98 &  Si I  & 5.87   &   --1.03  \\
 6437.64 &  Eu II & 1.32   &   --0.32  \\
 6471.66 &  Ca I  & 2.53   &   --0.69  \\
 6496.89 &  Ba II & 0.60   &   --0.41  \\
 6538.60 &   S I  & 8.05   &   --0.93  \\
 6645.13 &  Eu II & 1.38   &    0.12  \\
 6696.02 &  Al I  & 3.14   &   --1.57  \\
 6698.67 &  Al I  & 3.14   &   --1.87  \\
 6740.08 &  Nd II & 0.06   &   --1.53  \\
 6741.63 &  Si I  & 5.98   &   --1.65  \\
 6748.84 &   S I  & 7.87   &   --0.60  \\
 6757.17 &   S I  & 7.87   &   --0.31  \\
 6774.27 &  La II & 0.13   &   --1.71  \\
 6795.41 &   Y II & 1.74   &   --1.03  \\
 7034.90 &  Si I  & 5.87   &   --0.88  \\
 7375.25 &  Si I  & 6.19   &   --1.05  \\
 7423.50 &  Si I  & 5.62   &   --0.18  \\
 7680.27 &  Si I  & 5.86   &   --0.69  \\
 7849.97 &  Si I  & 6.19   &   --0.72  \\
 7881.88 &   Y II & 1.84   &   --0.57  \\
 7932.35 &  Si I  & 5.96   &   --0.47  \\
 7944.00 &  Si I  & 5.98   &   --0.31  \\
 8712.69 &  Mg I  & 5.93   &   --1.37  \\
 8736.02 &  Mg I  & 5.95   &   --0.36  \\
 8772.87 &  Al I  & 4.02   &   --0.17  \\
 8773.90 &  Al I  & 4.02   &   --0.02  \\
 9237.54 &   S I  & 6.53   &    0.04  \\
 9276.97 &  Zr I  & 0.69   &   --1.00  \\
\hline
\end{tabular}
\caption{List of the absorption lines used for the determination of the abundances. The list is ordered by wavelength and also displays the excitation potential ($\chi_{ex}$) and the logarithm of the weighted oscillator strength ($log~gf$) of the transition.} 
\label{linelist}
\end{table}

\newpage

\onecolumn
\section{Individual abundance results}

\begin{table}[!ht]
\centering
\scriptsize
\begin{tabular}{rrrrrrrrrrrrrrrrrrrrrr}     
\hline\hline
\multicolumn{1}{c}{Star} &  \multicolumn{1}{c}{Na} &  \multicolumn{1}{c}{Mg} &  \multicolumn{1}{c}{Al} &  \multicolumn{1}{c}{Si} &  \multicolumn{1}{c}{S}  &  \multicolumn{1}{c}{Ca} &  \multicolumn{1}{c}{Fe} \\
                         & \multicolumn{1}{c}{dex} & \multicolumn{1}{c}{dex} & \multicolumn{1}{c}{dex} & \multicolumn{1}{c}{dex} & \multicolumn{1}{c}{dex} & \multicolumn{1}{c}{dex} & \multicolumn{1}{c}{dex} \\
\hline
AAGem   &                       & -0.26 $\pm$ 0.00  (1) &                       & -0.24 $\pm$ 0.10  (11) & -0.24 $\pm$ 0.00  (1) & -0.37 $\pm$ 0.14  (2) & -0.35 $\pm$ 0.08 ~~(75) \\ 
ADGem   & -0.05 $\pm$ 0.00  (1) & -0.05 $\pm$ 0.05  (2) & -0.08 $\pm$ 0.09  (4) & -0.11 $\pm$ 0.15  (14) & -0.19 $\pm$ 0.00  (1) & -0.11 $\pm$ 0.14  (6) & -0.19 $\pm$ 0.13 (119) \\ 
ADPup   &  0.04 $\pm$ 0.13  (2) &  0.02 $\pm$ 0.12  (2) & -0.09 $\pm$ 0.18  (6) & -0.10 $\pm$ 0.09  (13) &  0.05 $\pm$ 0.00  (1) & -0.02 $\pm$ 0.09  (3) & -0.20 $\pm$ 0.12 (114) \\ 
AHVel   &  0.53 $\pm$ 0.05  (x) &  0.14 $\pm$ 0.10  (x) &  0.16 $\pm$ 0.12  (x) &  0.12 $\pm$ 0.15 ~~(x) &  0.24 $\pm$ 0.09  (x) &  0.04 $\pm$ 0.12  (x) & -0.04 $\pm$ 0.06 ~~~~(x) \\ 
AOAur   & -0.10 $\pm$ 0.00  (1) & -0.08 $\pm$ 0.08  (3) & -0.11 $\pm$ 0.11  (4) & -0.15 $\pm$ 0.16  (11) & -0.22 $\pm$ 0.00  (1) & -0.18 $\pm$ 0.08  (5) & -0.41 $\pm$ 0.11 (106) \\ 
AOCMa   &  0.25 $\pm$ 0.00  (1) &  0.27 $\pm$ 0.00  (1) &  0.35 $\pm$ 0.00  (1) &  0.24 $\pm$ 0.15 ~~(5) &  0.44 $\pm$ 0.11  (2) & -0.01 $\pm$ 0.26  (5) & -0.04 $\pm$ 0.13 ~~(35) \\ 
APPup   &  0.06 $\pm$ 0.01  (x) & -0.05 $\pm$ 0.05  (x) & -0.11 $\pm$ 0.11  (x) & -0.09 $\pm$ 0.12 ~~(x) &  0.23 $\pm$ 0.15  (x) & -0.14 $\pm$ 0.12  (x) & -0.15 $\pm$ 0.08 ~~~~(x) \\ 
AQCar   &                       &  0.38 $\pm$ 0.00  (1) &  0.28 $\pm$ 0.10  (3) &  0.18 $\pm$ 0.19 ~~(4) &                       &  0.36 $\pm$ 0.11  (2) & -0.30 $\pm$ 0.10 (106) \\ 
AQPup   &  0.12 $\pm$ 0.00  (1) &                       & -0.02 $\pm$ 0.18  (5) & -0.01 $\pm$ 0.10  (11) & -0.06 $\pm$ 0.28  (2) & -0.13 $\pm$ 0.14  (3) & -0.26 $\pm$ 0.03 ~~(89) \\ 
ATPup   &  0.29 $\pm$ 0.10  (2) &  0.19 $\pm$ 0.13  (2) & -0.20 $\pm$ 0.03  (2) & -0.04 $\pm$ 0.18  (11) &                       & -0.15 $\pm$ 0.17  (5) & -0.22 $\pm$ 0.10 ~~(75) \\ 
AVSgr   &  0.09 $\pm$ 0.00  (1) &  0.73 $\pm$ 0.00  (1) &  0.59 $\pm$ 0.00  (1) &  0.41 $\pm$ 0.12 ~~(7) &  0.75 $\pm$ 0.20  (2) &  0.43 $\pm$ 0.08  (3) &  0.26 $\pm$ 0.12 ~~~~(x) \\ 
AVTau   &  0.17 $\pm$ 0.00  (1) &  0.12 $\pm$ 0.09  (2) &  0.09 $\pm$ 0.23  (3) &  0.04 $\pm$ 0.18  (10) &  0.05 $\pm$ 0.00  (1) & -0.03 $\pm$ 0.09  (5) & -0.17 $\pm$ 0.11 ~~(86) \\ 
AXAur   &  0.17 $\pm$ 0.00  (1) & -0.01 $\pm$ 0.16  (2) & -0.09 $\pm$ 0.08  (2) & -0.02 $\pm$ 0.16  (13) & -0.08 $\pm$ 0.00  (1) & -0.04 $\pm$ 0.16  (6) & -0.22 $\pm$ 0.10 (109) \\ 
AXVel   &                       &  0.13 $\pm$ 0.00  (1) &  0.12 $\pm$ 0.06  (2) &  0.20 $\pm$ 0.21 ~~(7) &  0.38 $\pm$ 0.16  (2) &  0.03 $\pm$ 0.08  (4) & -0.15 $\pm$ 0.07 ~~(97) \\ 
BEMon   &  0.05 $\pm$ 0.00  (1) & -0.09 $\pm$ 0.05  (2) &  0.05 $\pm$ 0.12  (4) & -0.05 $\pm$ 0.14  (12) & -0.03 $\pm$ 0.00  (1) & -0.12 $\pm$ 0.12  (5) & -0.07 $\pm$ 0.09 (106) \\ 
BGVel   &                       &                       & -0.11 $\pm$ 0.06  (2) &  0.09 $\pm$ 0.07  (12) &  0.25 $\pm$ 0.00  (1) &  0.08 $\pm$ 0.14  (6) & -0.10 $\pm$ 0.08 (130) \\ 
BKAur   &                       &  0.32 $\pm$ 0.00  (x) & -0.03 $\pm$ 0.04  (x) &  0.16 $\pm$ 0.15 ~~(x) &  0.41 $\pm$ 0.09  (x) &  0.05 $\pm$ 0.14  (x) & -0.07 $\pm$ 0.11 ~~~~(x) \\ 
BNPup   &  0.32 $\pm$ 0.08  (2) &  0.16 $\pm$ 0.16  (2) &  0.09 $\pm$ 0.20  (5) &  0.12 $\pm$ 0.18  (12) &  0.21 $\pm$ 0.00  (1) &  0.09 $\pm$ 0.19  (5) & -0.03 $\pm$ 0.10 (128) \\ 
BVMon   &                       &  0.14 $\pm$ 0.11  (2) &  0.02 $\pm$ 0.18  (3) &  0.01 $\pm$ 0.20  (12) &                       & -0.03 $\pm$ 0.13  (5) & -0.10 $\pm$ 0.10 ~~(60) \\ 
CSOri   & -0.03 $\pm$ 0.00  (1) & -0.18 $\pm$ 0.04  (2) & -0.09 $\pm$ 0.00  (1) & -0.20 $\pm$ 0.22 ~~(5) & -0.29 $\pm$ 0.00  (1) & -0.20 $\pm$ 0.09  (4) & -0.19 $\pm$ 0.13 ~~(84) \\ 
CVMon   &  0.19 $\pm$ 0.00  (1) &  0.18 $\pm$ 0.01  (2) &  0.02 $\pm$ 0.09  (4) &  0.07 $\pm$ 0.17  (13) &  0.05 $\pm$ 0.00  (1) & -0.12 $\pm$ 0.13  (3) & -0.10 $\pm$ 0.13 (111) \\ 
DRVel   &  0.29 $\pm$ 0.00  (1) &  0.21 $\pm$ 0.13  (2) &  0.26 $\pm$ 0.12  (3) &  0.20 $\pm$ 0.13  (11) &  0.36 $\pm$ 0.23  (2) &  0.26 $\pm$ 0.20  (5) & -0.01 $\pm$ 0.07 (122) \\ 
EKMon   &                       &  0.05 $\pm$ 0.18  (2) &  0.02 $\pm$ 0.18  (4) & -0.06 $\pm$ 0.13  (11) & -0.15 $\pm$ 0.00  (1) & -0.04 $\pm$ 0.08  (5) & -0.05 $\pm$ 0.11 ~~(88) \\ 
EZVel   &  0.10 $\pm$ 0.00  (x) &  0.32 $\pm$ 0.00  (x) &  0.27 $\pm$ 0.12  (x) &  0.02 $\pm$ 0.15 ~~(x) &  0.13 $\pm$ 0.17  (x) &  0.12 $\pm$ 0.21  (x) & -0.01 $\pm$ 0.11 ~~~~(x) \\ 
HWPup   &                       & -0.02 $\pm$ 0.00  (1) & -0.07 $\pm$ 0.00  (1) & -0.38 $\pm$ 0.10 ~~(8) & -0.06 $\pm$ 0.42  (2) & -0.23 $\pm$ 0.08  (7) & -0.28 $\pm$ 0.18 ~~(35) \\ 
lCar    &                       &                       &  0.11 $\pm$ 0.13  (x) &  0.20 $\pm$ 0.11 ~~(x) &                       & -0.00 $\pm$ 0.00  (x) &  0.10 $\pm$ 0.20 ~~~~(x) \\ 
MYPup   &  0.13 $\pm$ 0.01  (2) & -0.18 $\pm$ 0.01  (2) & -0.18 $\pm$ 0.08  (3) & -0.06 $\pm$ 0.08  (15) & -0.07 $\pm$ 0.11  (2) & -0.14 $\pm$ 0.07  (7) & -0.25 $\pm$ 0.08 (113) \\ 
RSOri   &  0.40 $\pm$ 0.00  (1) &  0.10 $\pm$ 0.04  (2) &  0.16 $\pm$ 0.09  (3) &  0.12 $\pm$ 0.10 ~~(8) &  0.29 $\pm$ 0.14  (2) & -0.06 $\pm$ 0.25  (3) & -0.14 $\pm$ 0.13 (105) \\ 
RSPup   &  0.73 $\pm$ 0.00  (1) &  0.21 $\pm$ 0.00  (1) &  0.19 $\pm$ 0.06  (2) &  0.31 $\pm$ 0.08  (10) &  0.28 $\pm$ 0.01  (2) & -0.03 $\pm$ 0.00  (1) &  0.07 $\pm$ 0.09 (104) \\ 
RYCMa   &  0.12 $\pm$ 0.00  (x) &  0.07 $\pm$ 0.00  (x) &  0.12 $\pm$ 0.16  (x) &  0.07 $\pm$ 0.11 ~~(x) &  0.58 $\pm$ 0.00  (x) &  0.02 $\pm$ 0.10  (x) & -0.16 $\pm$ 0.09 ~~~~(x) \\ 
RYVel   &  0.44 $\pm$ 0.00  (1) &                       &  0.40 $\pm$ 0.08  (3) &  0.18 $\pm$ 0.12 ~~(9) &  0.48 $\pm$ 0.06  (2) &  0.13 $\pm$ 0.10  (3) & -0.05 $\pm$ 0.08 ~~(55) \\ 
RZCMa   &  0.06 $\pm$ 0.00  (1) &  0.12 $\pm$ 0.00  (1) &                       &                        &  0.16 $\pm$ 0.00  (1) &  0.12 $\pm$ 0.24  (6) & -0.20 $\pm$ 0.07 ~~(95)  \\ 
RZGem   &                       &                       &  0.04 $\pm$ 0.18  (4) & -0.14 $\pm$ 0.20  (11) & -0.29 $\pm$ 0.00  (1) & -0.22 $\pm$ 0.05  (4) & -0.44 $\pm$ 0.07 ~~(76) \\ 
RZVel   &                       &  0.43 $\pm$ 0.00  (1) &                       &  0.28 $\pm$ 0.12 ~~(8) &                       & -0.23 $\pm$ 0.25  (2) &  0.05 $\pm$ 0.10 ~~(77) \\ 
STTau   &                       &  0.08 $\pm$ 0.00  (1) &  0.11 $\pm$ 0.14  (5) &  0.08 $\pm$ 0.18 ~~(9) &  0.17 $\pm$ 0.00  (1) &  0.06 $\pm$ 0.02  (3) & -0.14 $\pm$ 0.10 ~~(82) \\ 
STVel   &                       &  0.26 $\pm$ 0.00  (1) &  0.09 $\pm$ 0.13  (4) &  0.12 $\pm$ 0.11  (12) &                       &  0.09 $\pm$ 0.10  (4) & -0.14 $\pm$ 0.10 (128) \\ 
SVMon   &                       & -0.08 $\pm$ 0.00  (1) &  0.12 $\pm$ 0.35  (2) & -0.02 $\pm$ 0.10 ~~(6) &                       & -0.27 $\pm$ 0.00  (1) & -0.10 $\pm$ 0.14 ~~(86) \\ 
SWVel   &  0.21 $\pm$ 0.00  (1) &  0.28 $\pm$ 0.00  (1) &  0.28 $\pm$ 0.09  (2) & -0.03 $\pm$ 0.10 ~~(4) &                       &  0.05 $\pm$ 0.00  (1) & -0.15 $\pm$ 0.08 ~~(66) \\ 
SXVel   &  0.32 $\pm$ 0.00  (1) &                       & -0.10 $\pm$ 0.11  (3) &  0.13 $\pm$ 0.14  (13) &  0.39 $\pm$ 0.00  (1) & -0.01 $\pm$ 0.15  (6) & -0.18 $\pm$ 0.07 (114) \\ 
SYAur   &  0.35 $\pm$ 0.00  (1) &  0.16 $\pm$ 0.23  (2) &  0.01 $\pm$ 0.03  (3) &  0.12 $\pm$ 0.12  (10) &  0.23 $\pm$ 0.06  (2) &  0.06 $\pm$ 0.14  (4) & -0.13 $\pm$ 0.11 ~~(79) \\ 
TVel    &  0.33 $\pm$ 0.00  (1) &  0.32 $\pm$ 0.09  (2) &  0.29 $\pm$ 0.02  (2) &  0.28 $\pm$ 0.18  (13) &  0.25 $\pm$ 0.00  (1) &  0.29 $\pm$ 0.12  (6) & -0.02 $\pm$ 0.07 (144) \\ 
TWCMa   &                       & -0.23 $\pm$ 0.00  (1) &  0.07 $\pm$ 0.18  (2) & -0.31 $\pm$ 0.09 ~~(3) &                       & -0.40 $\pm$ 0.05  (3) & -0.51 $\pm$ 0.09 (112) \\ 
TWMon   &                       &  0.35 $\pm$ 0.00  (1) &  0.13 $\pm$ 0.00  (1) &  0.14 $\pm$ 0.19 ~~(5) &  0.05 $\pm$ 0.00  (1) &  0.34 $\pm$ 0.10  (2) & -0.15 $\pm$ 0.08 ~~(37) \\ 
TXMon   &  0.07 $\pm$ 0.00  (1) & -0.12 $\pm$ 0.08  (2) & -0.28 $\pm$ 0.30  (5) & -0.16 $\pm$ 0.10  (14) & -0.23 $\pm$ 0.08  (2) & -0.17 $\pm$ 0.11  (8) & -0.12 $\pm$ 0.10 ~~(99) \\ 
TYMon   &  0.04 $\pm$ 0.00  (1) & -0.09 $\pm$ 0.14  (2) & -0.05 $\pm$ 0.14  (4) & -0.21 $\pm$ 0.04  (14) & -0.21 $\pm$ 0.00  (1) & -0.11 $\pm$ 0.11  (5) & -0.15 $\pm$ 0.11 (104) \\ 
TZMon   &  0.10 $\pm$ 0.00  (1) &  0.19 $\pm$ 0.13  (2) & -0.06 $\pm$ 0.18  (6) & -0.01 $\pm$ 0.11  (13) &  0.09 $\pm$ 0.01  (2) &  0.22 $\pm$ 0.07  (4) & -0.04 $\pm$ 0.08 ~~(82) \\ 
UXCar   &                       & -0.06 $\pm$ 0.00  (1) &  0.32 $\pm$ 0.06  (2) &  0.05 $\pm$ 0.06 ~~(9) &  0.41 $\pm$ 0.01  (2) &  0.14 $\pm$ 0.10  (4) & -0.10 $\pm$ 0.07 (111) \\ 
UYMon   & -0.08 $\pm$ 0.05  (2) & -0.23 $\pm$ 0.06  (2) & -0.25 $\pm$ 0.18  (4) & -0.22 $\pm$ 0.15  (15) & -0.15 $\pm$ 0.11  (2) & -0.28 $\pm$ 0.15  (7) & -0.33 $\pm$ 0.08 ~~(87) \\ 
UZSct   &  0.86 $\pm$ 0.00  (1) &  0.40 $\pm$ 0.00  (1) &                       &  0.47 $\pm$ 0.12 ~~(5) &                       &  0.23 $\pm$ 0.08  (4) &  0.35 $\pm$ 0.17 ~~~~(x) \\ 
V340Ara &                       &                       &  0.47 $\pm$ 0.00  (1) &  0.44 $\pm$ 0.05 ~~(3) &  0.80 $\pm$ 0.00  (1) &                       &  0.38 $\pm$ 0.17 ~~~~(x) \\ 
V397Car &                       &                       &  0.21 $\pm$ 0.09  (2) &  0.20 $\pm$ 0.16 ~~(8) &  0.11 $\pm$ 0.00  (1) &  0.09 $\pm$ 0.12  (4) & -0.08 $\pm$ 0.09 (113) \\ 
V495Mon &                       &  0.23 $\pm$ 0.10  (2) &                       & -0.01 $\pm$ 0.17 ~~(3) & -0.04 $\pm$ 0.00  (1) &  0.31 $\pm$ 0.13  (4) & -0.17 $\pm$ 0.15 ~~(58) \\ 
V508Mon &  0.22 $\pm$ 0.00  (1) &  0.06 $\pm$ 0.09  (2) & -0.01 $\pm$ 0.13  (3) & -0.01 $\pm$ 0.17  (11) & -0.10 $\pm$ 0.00  (1) & -0.09 $\pm$ 0.18  (5) & -0.25 $\pm$ 0.07 ~~(76) \\ 
V510Mon &  0.17 $\pm$ 0.00  (1) &  0.31 $\pm$ 0.00  (1) &  0.45 $\pm$ 0.00  (1) & -0.02 $\pm$ 0.13 ~~(7) &  0.18 $\pm$ 0.20  (2) &  0.02 $\pm$ 0.21  (3) & -0.12 $\pm$ 0.13 ~~(60) \\ 
VCar    &                       &  0.31 $\pm$ 0.00  (1) &  0.23 $\pm$ 0.16  (5) &  0.20 $\pm$ 0.12  (12) &  0.26 $\pm$ 0.00  (1) &  0.31 $\pm$ 0.13  (3) & -0.06 $\pm$ 0.07 (114) \\ 
VVel    & -0.01 $\pm$ 0.00  (1) &  0.05 $\pm$ 0.04  (2) & -0.06 $\pm$ 0.24  (4) & -0.12 $\pm$ 0.10  (10) & -0.19 $\pm$ 0.00  (1) & -0.11 $\pm$ 0.10  (5) & -0.30 $\pm$ 0.06 (113) \\ 
VXPup   &                       & -0.11 $\pm$ 0.03  (2) & -0.16 $\pm$ 0.14  (6) & -0.12 $\pm$ 0.17  (11) &  0.06 $\pm$ 0.13  (3) & -0.10 $\pm$ 0.12  (5) & -0.15 $\pm$ 0.12 (137) \\ 
VYCar   &  0.29 $\pm$ 0.00  (1) &  0.14 $\pm$ 0.00  (1) &                       &  0.49 $\pm$ 0.18 ~~(9) &  0.26 $\pm$ 0.00  (1) &  0.16 $\pm$ 0.10  (4) & -0.06 $\pm$ 0.06 ~~(91) \\ 
VYSgr   &  0.76 $\pm$ 0.13  (2) &  0.47 $\pm$ 0.00  (1) &  0.20 $\pm$ 0.12  (2) &  0.37 $\pm$ 0.05 ~~(6) &                       &  0.22 $\pm$ 0.10  (5) &  0.38 $\pm$ 0.17 ~~~~(x) \\ 
VZPup   &  0.06 $\pm$ 0.12  (2) & -0.09 $\pm$ 0.16  (2) & -0.12 $\pm$ 0.14  (5) & -0.21 $\pm$ 0.12 ~~(9) & -0.34 $\pm$ 0.00  (1) & -0.35 $\pm$ 0.14  (5) & -0.37 $\pm$ 0.07 (115) \\ 
WWMon   &  0.13 $\pm$ 0.00  (1) & -0.09 $\pm$ 0.24  (2) &                       & -0.02 $\pm$ 0.14 ~~(8) & -0.16 $\pm$ 0.08  (2) & -0.26 $\pm$ 0.11  (7) & -0.32 $\pm$ 0.13 ~~(48) \\ 
WXPup   &  0.22 $\pm$ 0.02  (2) &                       & -0.02 $\pm$ 0.10  (5) & -0.12 $\pm$ 0.14  (10) & -0.13 $\pm$ 0.28  (2) & -0.04 $\pm$ 0.08  (5) & -0.15 $\pm$ 0.09 (110) \\ 
XXMon   &  0.11 $\pm$ 0.00  (1) & -0.05 $\pm$ 0.11  (2) & -0.11 $\pm$ 0.17  (4) & -0.12 $\pm$ 0.13  (12) & -0.19 $\pm$ 0.22  (2) & -0.16 $\pm$ 0.14  (6) & -0.18 $\pm$ 0.11 ~~(99) \\ 
YAur    &  0.27 $\pm$ 0.00  (1) & -0.01 $\pm$ 0.21  (2) & -0.11 $\pm$ 0.20  (5) & -0.01 $\pm$ 0.20  (13) &  0.22 $\pm$ 0.02  (2) & -0.17 $\pm$ 0.10  (6) & -0.26 $\pm$ 0.12 ~~(88) \\ 
YZAur   &  0.22 $\pm$ 0.00  (1) & -0.20 $\pm$ 0.00  (1) &                       & -0.13 $\pm$ 0.12  (11) & -0.23 $\pm$ 0.00  (1) & -0.14 $\pm$ 0.10  (3) & -0.33 $\pm$ 0.12 ~~(68) \\ 
\hline
\end{tabular}
\caption{Light and $\alpha$ element abundance measurements for the Cepheids in our sample. The abundances are tabulated together with the rms uncertainties and with the number of lines used. (x) indicates that the final abundance is the mean value of the abundances derived from several spectra.}
\label{indiv_ab_light} 
\end{table}

\newpage

\begin{table}[!ht]
\centering
\scriptsize
\begin{tabular}{rrrrrrrrrrrrrrrrrrrrrr}
\hline\hline
\multicolumn{1}{c}{Star} &  \multicolumn{1}{c}{Y}  &  \multicolumn{1}{c}{Zr} &  \multicolumn{1}{c}{La} &  \multicolumn{1}{c}{Ce} &  \multicolumn{1}{c}{Nd} & \multicolumn{1}{c}{Eu} \\
                         & \multicolumn{1}{c}{dex} & \multicolumn{1}{c}{dex} & \multicolumn{1}{c}{dex} & \multicolumn{1}{c}{dex} & \multicolumn{1}{c}{dex} & \multicolumn{1}{c}{dex} \\
\hline                                                                                                       
AAGem   & -0.33 $\pm$ 0.10  (5) & -0.15 $\pm$ 0.00  (1) & -0.11 $\pm$ 0.05  (5) & -0.08 $\pm$ 0.08  (2) & -0.06 $\pm$ 0.14  (3) & -0.09 $\pm$ 0.06  (2) \\ 
ADGem   & -0.17 $\pm$ 0.06  (6) & -0.02 $\pm$ 0.00  (1) &  0.06 $\pm$ 0.05  (6) &  0.26 $\pm$ 0.06  (3) &  0.10 $\pm$ 0.07  (4) &  0.05 $\pm$ 0.04  (2) \\ 
ADPup   &  0.23 $\pm$ 0.21  (5) &  0.16 $\pm$ 0.00  (1) &  0.42 $\pm$ 0.26  (4) &  0.31 $\pm$ 0.13  (2) &  0.50 $\pm$ 0.19  (6) &  0.17 $\pm$ 0.13  (2) \\ 
AHVel   &  0.11 $\pm$ 0.09  (x) &  0.22 $\pm$ 0.00  (x) &  0.26 $\pm$ 0.15  (x) &  0.33 $\pm$ 0.07  (x) &  0.30 $\pm$ 0.16  (x) &  0.30 $\pm$ 0.00  (x) \\ 
AOAur   & -0.40 $\pm$ 0.11  (5) & -0.35 $\pm$ 0.00  (1) & -0.21 $\pm$ 0.17  (6) & -0.17 $\pm$ 0.23  (2) & -0.22 $\pm$ 0.23  (6) & -0.15 $\pm$ 0.00  (1) \\ 
AOCMa   &  0.03 $\pm$ 0.04  (2) &                       &  0.29 $\pm$ 0.13  (3) &                       &                       &                       \\ 
APPup   &  0.11 $\pm$ 0.09  (x) &  0.19 $\pm$ 0.00  (x) &  0.30 $\pm$ 0.12  (x) &  0.35 $\pm$ 0.05  (x) &  0.33 $\pm$ 0.14  (x) &  0.32 $\pm$ 0.09  (x) \\ 
AQCar   & -0.02 $\pm$ 0.15  (6) &                       &  0.02 $\pm$ 0.21  (4) &  0.02 $\pm$ 0.00  (1) &  0.25 $\pm$ 0.16  (2) &                       \\ 
AQPup   &  0.03 $\pm$ 0.05  (3) &                       &  0.08 $\pm$ 0.20  (4) &  0.09 $\pm$ 0.04  (2) &  0.02 $\pm$ 0.20  (3) &  0.03 $\pm$ 0.18  (2) \\ 
ATPup   & -0.19 $\pm$ 0.10  (5) & -0.12 $\pm$ 0.00  (1) & -0.11 $\pm$ 0.18  (4) &  0.06 $\pm$ 0.01  (2) &  0.06 $\pm$ 0.18  (4) & -0.21 $\pm$ 0.00  (2) \\ 
AVSgr   &  0.10 $\pm$ 0.02  (3) &                       &  0.05 $\pm$ 0.06  (4) &  0.12 $\pm$ 0.04  (2) &                       &  0.57 $\pm$ 0.16  (2) \\ 
AVTau   & -0.08 $\pm$ 0.09  (4) & -0.13 $\pm$ 0.00  (1) &  0.08 $\pm$ 0.08  (4) &                       &  0.28 $\pm$ 0.00  (1) &  0.16 $\pm$ 0.40  (2) \\ 
AXAur   & -0.35 $\pm$ 0.14  (6) & -0.22 $\pm$ 0.00  (1) & -0.22 $\pm$ 0.10  (6) & -0.08 $\pm$ 0.05  (2) & -0.10 $\pm$ 0.10  (4) & -0.31 $\pm$ 0.06  (2) \\ 
AXVel   &  0.02 $\pm$ 0.07  (4) &  0.03 $\pm$ 0.00  (1) &  0.28 $\pm$ 0.15  (4) &  0.44 $\pm$ 0.16  (2) &  0.16 $\pm$ 0.01  (2) &  0.14 $\pm$ 0.00  (1) \\ 
BEMon   &  0.06 $\pm$ 0.12  (4) & -0.04 $\pm$ 0.00  (1) &  0.11 $\pm$ 0.08  (6) &  0.11 $\pm$ 0.15  (4) &  0.05 $\pm$ 0.08  (4) &  0.03 $\pm$ 0.12  (2) \\ 
BGVel   &  0.08 $\pm$ 0.13  (6) &  0.17 $\pm$ 0.00  (1) &  0.24 $\pm$ 0.13  (3) &  0.35 $\pm$ 0.17  (3) &  0.27 $\pm$ 0.18  (4) &  0.21 $\pm$ 0.00  (1) \\ 
BKAur   &  0.08 $\pm$ 0.15  (x) &  0.14 $\pm$ 0.00  (x) &  0.26 $\pm$ 0.08  (x) &  0.31 $\pm$ 0.07  (x) &  0.28 $\pm$ 0.11  (x) &  0.27 $\pm$ 0.02  (x) \\ 
BNPup   &  0.26 $\pm$ 0.16  (5) &                       &  0.39 $\pm$ 0.16  (4) &  0.43 $\pm$ 0.04  (2) &  0.52 $\pm$ 0.18  (5) &  0.28 $\pm$ 0.05  (2) \\ 
BVMon   & -0.12 $\pm$ 0.24  (2) &                       &  0.24 $\pm$ 0.13  (2) &                       &                       &  0.27 $\pm$ 0.00  (1) \\ 
CSOri   & -0.20 $\pm$ 0.14  (3) &  0.15 $\pm$ 0.00  (1) &                       &  0.14 $\pm$ 0.00  (1) &  0.14 $\pm$ 0.28  (2) &                       \\ 
CVMon   & -0.09 $\pm$ 0.07  (7) &  0.04 $\pm$ 0.00  (1) &  0.06 $\pm$ 0.03  (4) &  0.08 $\pm$ 0.14  (4) &  0.07 $\pm$ 0.12  (4) &  0.03 $\pm$ 0.04  (2) \\ 
DRVel   &  0.13 $\pm$ 0.10  (5) &  0.40 $\pm$ 0.00  (1) &  0.14 $\pm$ 0.23  (5) &  0.29 $\pm$ 0.28  (2) &  0.17 $\pm$ 0.33  (4) & -0.04 $\pm$ 0.00  (1) \\ 
EKMon   & -0.19 $\pm$ 0.11  (4) &  0.12 $\pm$ 0.00  (1) & -0.05 $\pm$ 0.08  (4) &  0.18 $\pm$ 0.16  (3) &  0.16 $\pm$ 0.06  (4) &  0.01 $\pm$ 0.00  (1) \\ 
EZVel   &  0.20 $\pm$ 0.21  (x) &                       &  0.51 $\pm$ 0.15  (x) &  0.47 $\pm$ 0.09  (x) &  0.12 $\pm$ 0.22  (x) &  0.61 $\pm$ 0.28  (x) \\ 
HWPup   & -0.42 $\pm$ 0.18  (5) & -0.36 $\pm$ 0.00  (1) & -0.21 $\pm$ 0.17  (5) & -0.04 $\pm$ 0.00  (1) & -0.20 $\pm$ 0.11  (4) & -0.13 $\pm$ 0.13  (2) \\ 
lCar    &  0.29 $\pm$ 0.07  (x) &                       &  0.29 $\pm$ 0.19  (x) &  0.42 $\pm$ 0.13  (x) &  0.47 $\pm$ 0.16  (x) &  0.61 $\pm$ 0.00  (x) \\ 
MYPup   &  0.01 $\pm$ 0.10  (6) &  0.07 $\pm$ 0.00  (1) &  0.17 $\pm$ 0.13  (6) &  0.25 $\pm$ 0.15  (4) &  0.16 $\pm$ 0.19  (4) &  0.16 $\pm$ 0.08  (2) \\ 
RSOri   &  0.13 $\pm$ 0.16  (5) &  0.23 $\pm$ 0.00  (1) &  0.30 $\pm$ 0.14  (5) &  0.39 $\pm$ 0.00  (1) &  0.36 $\pm$ 0.23  (5) &  0.31 $\pm$ 0.06  (2) \\ 
RSPup   &  0.30 $\pm$ 0.20  (5) &                       &  0.24 $\pm$ 0.16  (4) &  0.19 $\pm$ 0.09  (2) &  0.39 $\pm$ 0.32  (4) &  0.37 $\pm$ 0.16  (2) \\ 
RYCMa   & -0.08 $\pm$ 0.12  (x) &  0.14 $\pm$ 0.00  (x) &  0.13 $\pm$ 0.16  (x) &  0.20 $\pm$ 0.07  (x) &  0.19 $\pm$ 0.19  (x) &  0.28 $\pm$ 0.00  (x) \\ 
RYVel   &  0.19 $\pm$ 0.22  (4) &                       &  0.07 $\pm$ 0.37  (4) &  0.15 $\pm$ 0.28  (2) &  0.17 $\pm$ 0.25  (5) &  0.50 $\pm$ 0.20  (2) \\ 
RZCMa   & -0.16 $\pm$ 0.16  (6) & -0.12 $\pm$ 0.00  (1) & -0.07 $\pm$ 0.16  (4) &  0.09 $\pm$ 0.04  (3) &  0.14 $\pm$ 0.09  (4) &  0.09 $\pm$ 0.00  (1) \\ 
RZGem   & -0.41 $\pm$ 0.19  (6) & -0.36 $\pm$ 0.00  (1) & -0.27 $\pm$ 0.06  (5) & -0.02 $\pm$ 0.15  (3) & -0.19 $\pm$ 0.08  (4) & -0.27 $\pm$ 0.04  (2) \\ 
RZVel   &  0.34 $\pm$ 0.19  (3) &                       &  0.27 $\pm$ 0.24  (3) &  0.25 $\pm$ 0.12  (2) &  0.54 $\pm$ 0.00  (1) &  0.37 $\pm$ 0.00  (1) \\ 
STTau   & -0.01 $\pm$ 0.08  (5) &  0.12 $\pm$ 0.00  (1) &  0.09 $\pm$ 0.10  (5) &  0.37 $\pm$ 0.11  (4) &  0.19 $\pm$ 0.16  (3) &  0.12 $\pm$ 0.06  (2) \\ 
STVel   & -0.04 $\pm$ 0.08  (7) &                       &  0.12 $\pm$ 0.12  (4) &  0.17 $\pm$ 0.00  (1) &  0.16 $\pm$ 0.19  (3) &  0.20 $\pm$ 0.00  (1) \\ 
SVMon   &  0.03 $\pm$ 0.15  (3) &  0.36 $\pm$ 0.00  (1) &  0.22 $\pm$ 0.02  (3) &  0.12 $\pm$ 0.16  (2) &  0.20 $\pm$ 0.16  (2) &  0.18 $\pm$ 0.00  (1) \\ 
SWVel   & -0.03 $\pm$ 0.00  (2) &                       &  0.10 $\pm$ 0.18  (5) &  0.04 $\pm$ 0.00  (1) & -0.02 $\pm$ 0.00  (1) &  0.08 $\pm$ 0.00  (1) \\ 
SXVel   &  0.01 $\pm$ 0.15  (6) &  0.17 $\pm$ 0.00  (1) &  0.15 $\pm$ 0.09  (5) &  0.19 $\pm$ 0.02  (2) &  0.05 $\pm$ 0.11  (3) &  0.19 $\pm$ 0.00  (1) \\ 
SYAur   & -0.07 $\pm$ 0.17  (6) &                       &  0.04 $\pm$ 0.13  (4) &  0.22 $\pm$ 0.05  (2) &  0.17 $\pm$ 0.11  (4) &  0.14 $\pm$ 0.10  (2) \\ 
TVel    &  0.30 $\pm$ 0.05  (5) &  0.42 $\pm$ 0.00  (1) &  0.49 $\pm$ 0.12  (5) &  0.63 $\pm$ 0.20  (4) &  0.39 $\pm$ 0.16  (4) &  0.45 $\pm$ 0.00  (1) \\ 
TWCMa   & -0.32 $\pm$ 0.07  (6) &                       & -0.03 $\pm$ 0.18  (4) &                       & -0.10 $\pm$ 0.07  (4) &  0.05 $\pm$ 0.00  (1) \\ 
TWMon   &  0.21 $\pm$ 0.07  (2) &                       &  0.20 $\pm$ 0.07  (4) &                       &                       &                       \\ 
TXMon   & -0.12 $\pm$ 0.09  (6) & -0.02 $\pm$ 0.00  (1) &  0.09 $\pm$ 0.11  (6) &  0.04 $\pm$ 0.09  (4) &  0.04 $\pm$ 0.07  (4) & -0.02 $\pm$ 0.01  (2) \\ 
TYMon   & -0.11 $\pm$ 0.13  (6) & -0.05 $\pm$ 0.00  (1) & -0.05 $\pm$ 0.17  (5) &  0.02 $\pm$ 0.00  (1) &  0.07 $\pm$ 0.12  (5) &  0.09 $\pm$ 0.01  (2) \\ 
TZMon   & -0.05 $\pm$ 0.17  (6) &  0.01 $\pm$ 0.00  (1) &  0.01 $\pm$ 0.08  (6) &  0.16 $\pm$ 0.19  (4) &  0.24 $\pm$ 0.25  (5) &  0.06 $\pm$ 0.16  (2) \\ 
UXCar   & -0.01 $\pm$ 0.10  (5) &  0.21 $\pm$ 0.00  (1) &  0.11 $\pm$ 0.04  (4) &  0.25 $\pm$ 0.09  (2) &  0.18 $\pm$ 0.21  (3) &  0.14 $\pm$ 0.00  (1) \\ 
UYMon   & -0.35 $\pm$ 0.11  (6) & -0.15 $\pm$ 0.00  (1) & -0.12 $\pm$ 0.05  (6) & -0.04 $\pm$ 0.06  (3) &  0.01 $\pm$ 0.16  (5) & -0.11 $\pm$ 0.06  (2) \\ 
UZSct   &  0.37 $\pm$ 0.25  (2) &                       &  0.48 $\pm$ 0.10  (3) &  0.57 $\pm$ 0.16  (2) &                       &  0.66 $\pm$ 0.00  (1) \\ 
V340Ara &                       &                       &  0.22 $\pm$ 0.12  (2) &                       &                       &  0.32 $\pm$ 0.00  (1) \\ 
V397Car & -0.04 $\pm$ 0.06  (4) &                       &  0.06 $\pm$ 0.13  (4) &  0.06 $\pm$ 0.00  (1) &  0.02 $\pm$ 0.11  (2) &  0.12 $\pm$ 0.00  (1) \\ 
V495Mon &                       &                       &                       &                       &  0.36 $\pm$ 0.00  (1) &  0.18 $\pm$ 0.00  (1) \\ 
V508Mon & -0.14 $\pm$ 0.13  (6) & -0.12 $\pm$ 0.00  (1) &  0.05 $\pm$ 0.02  (3) &  0.13 $\pm$ 0.13  (3) &  0.02 $\pm$ 0.12  (3) &  0.15 $\pm$ 0.00  (1) \\ 
V510Mon & -0.17 $\pm$ 0.23  (3) &                       &                       &  0.33 $\pm$ 0.12  (2) &                       &  0.02 $\pm$ 0.04  (2) \\ 
VCar    &  0.07 $\pm$ 0.12  (5) &  0.29 $\pm$ 0.00  (1) &  0.26 $\pm$ 0.17  (5) &  0.15 $\pm$ 0.25  (2) &  0.16 $\pm$ 0.27  (5) &  0.38 $\pm$ 0.00  (1) \\ 
VVel    & -0.13 $\pm$ 0.13  (7) & -0.02 $\pm$ 0.00  (1) &  0.07 $\pm$ 0.14  (5) &  0.24 $\pm$ 0.13  (3) &  0.19 $\pm$ 0.09  (4) & -0.04 $\pm$ 0.00  (1) \\ 
VXPup   &  0.04 $\pm$ 0.12  (4) &  0.02 $\pm$ 0.00  (1) &  0.32 $\pm$ 0.09  (6) &  0.35 $\pm$ 0.01  (3) &  0.26 $\pm$ 0.19  (5) &  0.23 $\pm$ 0.13  (2) \\ 
VYCar   &  0.20 $\pm$ 0.09  (5) &  0.34 $\pm$ 0.00  (1) &  0.32 $\pm$ 0.12  (5) &  0.39 $\pm$ 0.01  (2) &  0.38 $\pm$ 0.23  (2) &  0.39 $\pm$ 0.00  (1) \\ 
VYSgr   &  0.23 $\pm$ 0.15  (2) &                       &  0.42 $\pm$ 0.29  (3) &  0.31 $\pm$ 0.23  (2) &  0.13 $\pm$ 0.00  (1) &  0.50 $\pm$ 0.07  (2) \\ 
VZPup   & -0.08 $\pm$ 0.18  (4) &  0.10 $\pm$ 0.00  (1) &  0.07 $\pm$ 0.12  (4) &  0.14 $\pm$ 0.25  (2) & -0.09 $\pm$ 0.12  (4) & -0.04 $\pm$ 0.05  (2) \\ 
WWMon   & -0.26 $\pm$ 0.17  (4) &  0.02 $\pm$ 0.00  (1) & -0.23 $\pm$ 0.06  (2) & -0.21 $\pm$ 0.00  (1) & -0.14 $\pm$ 0.22  (2) & -0.02 $\pm$ 0.00  (1) \\ 
WXPup   & -0.01 $\pm$ 0.10  (4) &  0.21 $\pm$ 0.00  (1) &  0.13 $\pm$ 0.25  (4) &  0.18 $\pm$ 0.01  (2) &  0.24 $\pm$ 0.21  (5) & -0.05 $\pm$ 0.04  (2) \\ 
XXMon   & -0.23 $\pm$ 0.18  (5) & -0.10 $\pm$ 0.00  (1) &  0.05 $\pm$ 0.13  (3) &                       &  0.03 $\pm$ 0.15  (3) & -0.06 $\pm$ 0.06  (2) \\ 
YAur    & -0.36 $\pm$ 0.05  (6) & -0.30 $\pm$ 0.00  (1) & -0.24 $\pm$ 0.10  (2) & -0.21 $\pm$ 0.03  (2) & -0.24 $\pm$ 0.06  (3) & -0.11 $\pm$ 0.00  (1) \\ 
YZAur   & -0.34 $\pm$ 0.12  (4) & -0.31 $\pm$ 0.00  (1) & -0.19 $\pm$ 0.10  (5) &  0.25 $\pm$ 0.00  (1) &                       & -0.16 $\pm$ 0.00  (1) \\ 
\hline
\end{tabular}
\caption{Same as Table \ref{indiv_ab_light}, but for heavy elements} 
\label{indiv_ab_heavy}
\end{table}

\end{appendix}


\begin{thebibliography}{}
   \bibitem[Afflerbach et al.(1996)]{Aff1996} Afflerbach, A., Churchwell, E., Acord, J. M., Hofner, P., Kurtz, S., Depree, C. G.,
	ApJS 106, 423	
   \bibitem[Afflerbach et al.(1997)]{Aff1997} Afflerbach, A., Churchwell, E., Werner, M. W., 1997, 
        ApJ 478, 190
   \bibitem[Andrievsky et al.(2002a)]{And2002a} Andrievsky, S. M., Kovtyukh, V. V., Luck, R. E., L\'epine, J. R. D., Bersier, D., Maciel, W. J., Barbuy, B., Klochkova, V. G., Panchuk, V. E., Karpischek, R. U., 2002, 
        A\&A 381, 32
   \bibitem[Andrievsky et al.(2002b)]{And2002b} Andrievsky, S. M., Bersier, D., Kovtyukh, V. V., Luck, R. E., Maciel, W. J.,  L\'epine, J. R. D., Beletsky, Yu. V., 2002, 
        A\&A 384, 140
   \bibitem[Andrievsky et al.(2002c)]{And2002c} Andrievsky, S. M., Kovtyukh, V. V., Luck, R. E., L\'epine, J. R. D., Maciel, W. J., Beletsky, Yu. V., 2002, 
        A\&A 392, 491
   \bibitem[Andrievsky et al.(2004)]{And2004} Andrievsky, S. M., Luck, R. E., Martin, P., L\'epine, J. R. D., 2004, 
        A\&A 413, 159
   \bibitem[Andrievsky et al.(2005)]{And2005} Andrievsky, S. M., Luck, R. E., Kovtyukh, V. V., 2005, 
        AJ 130, 1880
   \bibitem[Andrievsky et al.(2013)]{And2013} Andrievsky, S. M., Lépine, J. R. D., Korotin, S. A., Luck, R. E., Kovtyukh, V. V., Maciel, W. J., 2013,
	MNRAS 428, 3252
   \bibitem[Balser et al.(2011)]{Balser2011} Balser, D. S., Rood, R. T., Bania, T. M., Anderson, L., 2011,
	ApJ 738, 27
   \bibitem[Barker et al.(2007)]{Bar2007} Barker, M. K., Sarajedini, A., Geisler, D., Harding, P., Schommer, R., 2007, 
        AJ 133, 1138
   \bibitem[Bilir et al.(2012)]{Bilir2012} Bilir, S., Karaali, S., Ak, S., \"{O}nal, \"{O}., Da\u{g}tekin, N. D., Yontan, T., Gilmore, G., Seabroke, G. M., 2012,
	MNRAS 421, 3362	
   \bibitem[Binney \& Tremaine(2008)]{Bin2008} Binney, J., Tremaine, S., 2008, 
	Galactic Dynamics: Second Edition
   \bibitem[Bird et al.(2012)]{Bird2012} Bird, J. C., Kazantzidis, S., Weinberg, D. H., 2012,
	MNRAS 420, 913
   \bibitem[Bird et al.(2012b)]{Bird2012b} Bird, J. C., Kazantzidis, S., Weinberg, D. H., Guedes, J., Callegari, S., Mayer, L., Madau, P., 2012
	{\it Arxiv:1301.0620}	
   \bibitem[Bono et al.(1999)]{Bono1999} Bono, G., Caputo, F., Castellani, V., Marconi, M., 1999, 
        ApJ 512, 711
   \bibitem[Bono et al.(2005)]{Bono2005} Bono, G., Marconi, M., Cassisi, S., Caputo, F., Gieren, W., Pietrzy\'nski, G., 2005
	ApJ 621, 966
   \bibitem[Bono et al.(2008)]{Bono2008} Bono, G., Caputo, F., Fiorentino, G., Marconi, M., Musella, I., 2008,
        ApJ 684, 102
   \bibitem[Bono et al.(2010)]{Bono2010} Bono, G., Caputo, F., Marconi, M., Musella, I., 2010,
        ApJ 715, 277
   \bibitem[Bragaglia et al.(2008)]{Bra2008} Bragaglia, A., Sestito, P., Villanova, S., Carretta, E., Randich, S., Tosi, M., 2008,
	A\&A 480, 79
   \bibitem[Bresolin et al.(2009a)]{Bre2009a} Bresolin, F., Ryan-Weber, E., Kennicutt, R. C., Goddard, Q., 2009,
        ApJ 695, 580
   \bibitem[Bresolin et al.(2009b)]{Bre2009b} Bresolin, F., Gieren, W., Kudritzki, R-P., Pietrzy\'nski, G., Urbaneja, M. A., Carraro, G., 2009,
        ApJ 700, 309
   \bibitem[Carraro et al.(2007)]{Carra2007} Carraro, G., Geisler, D., Villanova, S., Frinchaboy, P. M., Majewski, S. R., 2007, 
	A\&A 476, 217
   \bibitem[Carretta et al.(2004)]{Carre2004} Carretta, E., Bragaglia, A., Gratton, R. G., Tosi, M., 2004, 
	A\&A 422, 951
   \bibitem[Carretta et al.(2005)]{Carre2005} Carretta, E., Bragaglia, A., Gratton, R. G., Tosi, M., 2005,
	A\&A 441, 131
   \bibitem[Carretta et al.(2007)]{Carre2007} Carretta, E., Bragaglia, A., Gratton, R. G., 2007,
	A\&A 473, 129
   \bibitem[Cescutti et al.(2007)]{Ces2007} Cescutti, G., Matteucci, F., Fran\c cois, P., Chiappini, C., 2007, 
        A\&A 462, 943
   \bibitem[Charbonneau(1995)]{Char1995} Charbonneau, P., 1995, 
	ApJ SS 101, 309
   \bibitem[Chen et al.(2003)]{Chen2003} Chen, L., Hou, J. L., Wang, J. J., 2003, 
	AJ 125, 1397
   \bibitem[Cheng et al.(2012a)]{Cheng2012a} Cheng, J. Y., Rockosi, C. M., Morrison, H. L., Sch\"{o}nrich, R. A., Lee, Y. S., Beers, T. C., Bizyaev, D., Pan, K., Schneider, D. P., 2012,
	ApJ 746, 149
   \bibitem[Cheng et al.(2012b)]{Cheng2012b} Cheng, J. Y., Rockosi, C. M., Morrison, H. L., Lee, Y. S., Beers, T. C., Bizyaev, D., Harding, P., Malanushenko, E., Malanushenko, V., Oravetz, D., Pan, K., Schlesinger, K. J., Schneider, D. P., Simmons, A., Weaver, B. A., 2012,
	ApJ 752, 51
   \bibitem[Chiappini et al.(1997)]{Chia1997} Chiappini, C., Matteucci, F., Gratton, R., 1997, 
        ApJ 477, 765
   \bibitem[Chiappini et al.(2001)]{Chia2001} Chiappini, C., Matteucci, F., Romano, D., 2001, 
        ApJ 554, 1044
   \bibitem[Colavitti et al.(2008)]{Cola2008} Colavitti, E., Cescutti, G., Matteucci, F., Chiappini, C., 2008,
	A\&A 496, 429
   \bibitem[Co\c{s}kuno\u{g}lu et al.(2012)]{Cosku2012} Co\c{s}kuno\u{g}lu, B., Ak, S., Bilir, S., Karaali, S., \"{O}nal, \"{O}., Yaz, E., Gilmore, G., Seabroke, G. M., 2012, 
        MNRAS 419, 2844
   \bibitem[Costa et al.(2004)]{Cos2004} Costa, R. D. D., Uchida, M. M. M., Maciel, W. J., 2004, 
        A\&A 423, 199
   \bibitem[Daflon \& Cunha(2004b)]{Daf2004b} Daflon, S., Cunha, K., 2004, 
        ApJ 617, 1115
   \bibitem[Dambis(2009)]{Dam2009} Dambis, A. K., 2009,
        MNRAS 396, 553
   \bibitem[Deharveng et al.(2000)]{Deh2000} Deharveng, L., Pena, M., Caplan, J., Costero, R., 2000, 
	MNRAS 311, 329
   \bibitem[Denissenkov(1994)]{Deni1994} Denissenkov, P. A., 1994, 
	A\&A 287, 113
   \bibitem[Di Matteo et al.(2013)]{diMatt2013} Di Matteo, P., Haywood, M., Combes, F., Semelin, B., Snaith, O. N., 2013,
	A\&A 553, A102	
   \bibitem[Edvardsson et al.(1993)]{Edv1993} Edva	Steinmetz, M.; Zwitter, T.; Siebert, A.; Watson, F. G.; Freeman, K. C.; Munari, U.; Campbell, R.; Williams, M.; Seabroke, G. M.; Wyse, R. F. G.; 
rdsson, B., Andersen, J., Gustafsson, B., Lambert, D. L., Nissen, P. E., Tomkin, J., 1993, 
	A\&A 275, 101
   \bibitem[Esteban et al.(2005)]{Este2005} Esteban, C., García-Rojas, J., Peimbert, M., Peimbert, A., Ruiz, M. T., Rodríguez, M., Carigi, L., 2005,
	ApJ 618, 95
   \bibitem[Fernie et al.(1995)]{Fer1995} Fernie, J. D., Beattie, B., Evans, N. R., Seager, S., 1995, 
        IBVS 4148
   \bibitem[Few et al.(2012)]{Few2012} Few, C. G., Courty, S., Gibson, B. K., Kawata, D., Calura, F., Teyssier, R., 2012,
	MNRAS 424, 11
   \bibitem[Fiorentino et al.(2007)]{Fio2007} Fiorentino, G., Marconi, M., Musella, I., Caputo, F., 2007, 
        A\&A 476, 863
   \bibitem[Fouqu\' e et al.(2007)]{Fou2007} Fouqu\' e, P., Arriagada, P., Storm, J., Barnes, T. G., Nardetto, N., M\' erand, A., Kervella, P., Gieren, W., Bersier, D., Benedict, G. F., McArthur, B. E., 2007, 
        A\&A 476, 73
   \bibitem[Fran\c cois \& Matteucci(1993)]{Fran1993} Fran\c cois, P., Matteucci, F., 1993,
	A\&A 280, 136	
   \bibitem[Freeman(2012)]{Free2012} Freeman, K. C., 2012, 
	ASPC 458, 393
\bibitem[Friel et al.(2002)]{Frie2002} Friel, E. D., Janes, K. A., Tavarez, M., Scott, J., Katsanis, R., Lotz, J., Hong, L., Miller, N., 2002, 
        AJ 124, 2693
   \bibitem[Friel et al.(2010)]{Frie2010} Friel, E. D., Jacobson, H. R., Pilachowski, C. A., 2010,
	AJ 139, 1942
   \bibitem[Fritz et al.(2011)]{Fritz2011} Fritz, T. K., Gillessen, S., Dodds-Eden, K., Lutz, D., Genzel, R., Raab, W., Ott, T., Pfuhl, O., Eisenhauer, F., Yusef-Zadeh, F., 2011,
        ApJ 737, 73
   \bibitem[Fu et al.(2009)]{Fu2009} Fu, J., Hou, J. L., Yin, J., Chang, R. X., 2009,
	ApJ 696, 668
   \bibitem[Fu et al.(2013)]{Fu2013} Fu, J., Kauffmann, G., Huang, M., Yates, R. M., Moran, S., Heckman, T. M., Dav\' e, R., Guo, Q., 2013, 
arXiv1303.5586F	
   \bibitem[Gazzano et al.(2013)]{Gazza2013} Gazzano, J.-C., Kordopatis, G., Deleuil, M., de Laverny, P., Recio-Blanco, A., Hill, V., 2013,
	A\&A 550, A125
   \bibitem[Genovali et al.(2013a)]{Geno2013a} Genovali, K., Lemasle, B., Bono, G., Romaniello, M., Primas, F., Fabrizio, M., Buonanno, R., Fran\c cois, P., Inno, L., Laney, C. D., Matsunaga, N., Pedicelli, S., Thévenin, F., 2013a,
	A\&A 554, A132
   \bibitem[Genovali et al.(2013b)]{Geno2013b} Genovali, K., et al, 2013b, in prep.
   \bibitem[Gerke et al.(2011)]{Ger2011} Gerke, J. R., Kochanek, C. S., Prieto, J. L., Stanek, K. Z., Macri, L. M., 2011,
        ApJ 743, 176
   \bibitem[Ghez et al.(2008)]{Ghez2008} Ghez, A. M., Salim, S., Weinberg, N. N., Lu, J. R., Do, T., Dunn, J. K., Matthews, K., Morris, M. R., Yelda, S., Becklin, E. E., Kremenek, T., Milosavljevic, M., Naiman, J., 2008,
        ApJ 689, 1044
   \bibitem[Gillessen et al.(2009)]{Gill2009} Gillessen, S., Eisenhauer, F., Trippe, S., Alexander, T., Genzel, R., Martins, F., Ott, T., 2009,
        ApJ 692, 1075
   \bibitem[Gilmore et al.(2012)]{Gil2012} Gilmore, G., Randich, S., Asplund, M. et al., 2012, 
	The Messenger 147, 25
   \bibitem[Grevesse et al.(1996)]{Gre1996} Grevesse, N., Noels, A., Sauval, J., 1996, 
	ASPC 99, 117
   \bibitem[Groenewegen et al.(2008)]{Groe2008} Groenewegen, M. A. T., Udalski, A., Bono, G., 2008, 
        A\&A 481, 441   
   \bibitem[Gummersbach et al.(1998)]{Gum1998} Gummersbach, C. A., Kaufer, D. R., Sch\"afer, D. R., Szeifert,T., Wolf, B., 1998, 
	A\&A, 338, 881
   \bibitem[Harris(1981)]{Har1981} Harris, H. C., 1981, 
        AJ 86, 707
   \bibitem[Harris(1984)]{Har1984} Harris, H. C., 1984, 
        ApJ 282, 655
   \bibitem[Henry et al.(2004)]{Hen2004} Henry, R. B. C., Kwitter, K. B., Balick, B., 2004,
	AJ 127, 2284
   \bibitem[Henry et al.(2010)]{Hen2010} Henry, R. B. C., Kwitter, K. B., Jaskot, A. E., Balick, B., Morrison, M., Milingo, J. B., 2010,
	ApJ 724, 748
   \bibitem[Inno et al.(2013)]{Inno2013} Inno, L., Matsunaga, N., Bono, G., Caputo, F., Buonanno, R., Genovali, K., Laney, C. D., Marconi, M., Piersimoni, A. M., Primas, F., Romaniello, M., 2013,
	ApJ 764, 84
   \bibitem[Jacobson et al.(2008)]{Jaco2008} Jacobson, H. R., Friel, E. D., Pilachowski, C. A. 2008, 
	AJ 135, 2341
   \bibitem[Jacobson et al.(2009)]{Jaco2009} Jacobson, H. R., Friel, E. D., Pilachowski, C. A. 2009, 
	AJ 137, 4753
   \bibitem[Jacobson et al.(2011a)]{Jaco2011a} Jacobson, H. R., Friel, E. D., Pilachowski, C. A. 2011a, 
	AJ 141, 58
   \bibitem[Jacobson et al.(2011b)]{Jaco2011b} Jacobson, H. R., Friel, E. D., Pilachowski, C. A. 2011b, 
	AJ 142, 59
   \bibitem[Kobayashi \& Nakasato(2011)]{Koba2011} Kobayashi, C., Nakasato, N., 2011, 
	ApJ 729, 16
   \bibitem[Kovtyukh \& Andrievsky(1999)]{Kov1999} Kovtyukh, V. V., Andrievsky, S. M., 1999,
	A\&A 351, 597	
   \bibitem[Kovtyukh \& Gorlova(2000)]{KovGor2000} Kovtyukh, V. V., Gorlova, N. I., 2000, 
	A\&A 358, 587
   \bibitem[Kovtyukh et al.(2005a)]{Kov2005a} Kovtyukh, V. V., Andrievsky, S. M.,  Belik, S. I., Luck, R. E., 2005, 
        AJ 129, 433
   \bibitem[Kovtyukh et al.(2005b)]{Kov2005b} Kovtyukh, V. V., Wallerstein, G., Andrievsky, S. M., 2005, 
        PASP 117, 1173
\bibitem[Kovtyukh(2007)]{Kov2007} Kovtyukh, V. V., 2007, 
        MNRAS 378, 617
   \bibitem[Kudritzki et al.(2008)]{Ku2008} Kudritzki, R-P., Urbaneja, M. A., Bresolin, F., Przybilla, N., Gieren, W., Pietrzy\'nski, G., 2008, 
        ApJ 681, 269
   \bibitem[Kupka et al.(1999)]{Kup1999} Kupka, F., Piskunov, N. E., Ryabchikova, T. A., Stempels, H. C., Weiss, W. W., 1999, 
	A\&AS, 138, 119
   \bibitem[Lacey \& Fall(1985)]{Lacey1985} Lacey, C. G., Fall, S. M., 1985,
	ApJ 290, 154	
   \bibitem[Laney \& Stobie(1994)]{LS1994} Laney, C. D., Stobie, R. S., 1994, 
        MNRAS 266, 441
   \bibitem[Laney \& Caldwell(2007)]{Laney2007} Laney, C. D., Caldwell, J. A. R., 2007,
        MNRAS 377, 147
   \bibitem[Lemasle et al.(2007)]{Lem2007} Lemasle, B., Fran\c cois, P., Bono, G., Mottini, M., Primas, F., Romaniello, M., 2007, 
        A\&A 467, 283
   \bibitem[Lemasle et al.(2008)]{Lem2008} Lemasle, B., Fran\c cois, P., Piersimoni, A. Pedicelli, S., Bono, G., Laney, C. D., Primas, F., Romaniello, M., 2008, 
	A\&A 490, 613
   \bibitem[Loeillet et al.(2008)]{Loei2008} Loeillet, B., Bouchy, F., Deleuil, M., Royer, F., Bouret, J. C., Moutou, C., Barge, P., de Laverny, P., Pont, F., Recio-Blanco, A., Santos, N. C., 2008,
	A\&A 479, 865
   \bibitem[Luck et al.(2003)]{Luck2003} Luck, R. E., Gieren, W. P., Andrievsky, S. M., Kovtyukh, V. V., Fouqu\'e, P., Pont, P., Kienzle, F., 2003, 
        A\&A 401, 939
   \bibitem[Luck \& Andrievsky(2004)]{Luck2004} Luck, R. E., Andrievsky, S. M., 2004, 
        AJ 128, 343 
   \bibitem[Luck et al.(2006)]{Luck2006} Luck, R. E., Kovtyukh, V. V., Andrievsky, S. M., 2006, 
        AJ 132, 902
   \bibitem[Luck et al.(2008)]{Luck2008} Luck, R. E., Andrievsky, S. M., Fokin, A., Kovtyukh, V. V., 2008, 
        AJ 136, 98
   \bibitem[Luck et al.(2011)]{Luck2011a} Luck, R. E.; Andrievsky, S. M., Kovtyukh, V. V., Gieren, W., Graczyk, D., 2011,
        AJ 142, 51
   \bibitem[Luck \& Lambert(2011)]{Luck2011b} Luck, R. E., Lambert, D. L.,  2011,
        AJ 142, 136
   \bibitem[Maciel et al.(1999)]{Mac1999} Maciel, W. J., Quireza, C., 1999, 
	A\&A 345, 629
   \bibitem[Maciel et al.(2003)]{Mac2003} Maciel, W. J., Costa, R. D. D., Uchida, M. M. M., 2003, 
	A\&A 397, 667
   \bibitem[Maciel \& Costa(2009)]{Mac2009} Maciel, W. J., Costa, R. D. D., 2009, 
	in IAU Symposium, Vol. 254, IAU Symposium, ed. J. Andersen, J. Bland-Hawthorn, \& B. Nordstr\"om, 38
   \bibitem[Madore(1982)]{Mado1982} Madore, B. F., 1982,
        ApJ 253, 575
   \bibitem[Madore \& Freedman(2009)]{Mado2009} Madore, B. F., Freedman, W. F., 2009,
        ApJ 696, 1498
   \bibitem[Magrini et al.(2009a)]{Mag2009a} Magrini, L., Sestito, P., Randich, S., Galli, D., 2009,
	A\&A 494, 95
   \bibitem[Magrini et al.(2009b)]{Mag2009b} Magrini, L., Stanghellini, L., Villaver, E., 2009,
        ApJ 696, 729
   \bibitem[Magrini et al.(2010)]{Mag2010} Magrini, L., Randich, S., Zoccali, M., Jilkova, L., Carraro, G., Galli, D., Maiorca, E., Busso, M., 2010,
	A\&A 523, A11
   \bibitem[Majaess(2010)]{Maja2010} Majaess, D., 2010,
        AcA 60, 55
   \bibitem[Majaess et al.(2011)]{Maja2011} Majaess, D., Turner, D., Gieren, W., 2011,-0.063
        ApJ 741, 36
   \bibitem[Majewski et al.(2007)]{Maje2007} Majewski, S. R., Skrutskie, M. F., Schiavon, R. P., Wilson, J. C., O'Connell, R. W., Smith, V. V., Shetrone, M., Cunha, K., Frinchaboy, P. M., Reid, I. N. et al., 2007, 
	BAAS 39, 962
\bibitem[Malkin(2012)]{Mal2012} Malkin, Z., 2012, 
	{\it arXiv:1202.6128}
   \bibitem[Matsunaga et al.(2009)]{Matsu2009} Matsunaga, N., Kawadu, T., Nishiyama, S., Nagayama, T., Hatano, H., Tamura, M., Glass, I. S., Nagata, T., 2009,
        MNRAS 399, 1709
   \bibitem[Matsunaga et al.(2013)]{Matsu2013} Matsunaga, N., Feast, M. W., Kawadu, T., Nishiyama, S., Nagayama, T., Nagata, T., Tamura, M., Bono, G., Kobayashi, N., 2013,
	MNRAS 429, 385
   \bibitem[Matteucci \& Fran\c cois(1989)]{Matt1989} Matteucci, F., Fran\c cois, P., 1989, 
        MNRAS 239, 885
   \bibitem[Merle et al.(2011)]{Merle2011} Merle, T., Th\'{e}venin, F., Pichon, B., Bigot, L., 2011,
	MNRAS 418, 863
   \bibitem[Minchev \& Famaey(2010)]{Min2010} Minchev, I., Famaey, B., Combes, F., Di Matteo P., Mouhcine, M., Wozniak, H., 2011, 
	A\&A 527, A147
   \bibitem[Minchev et al.(2011)]{Min2011} Minchev, I., Famaey, B., 2010, 
	ApJ 722, 112
   \bibitem[Minchev et al.(2012)]{Min2012} Minchev, I., Chiappini, C., Martig, M., 2012,
	{\it ArXiv:1208.1506}
   \bibitem[Moll\' a \& D\' iaz(2005)]{Molla2005} Moll\' a, M., D\' iaz, A. I., 2005,
	MNRAS 358, 521
   \bibitem[Monson \& Pierce(2011)]{Mon2011} Monson, A. J., Pierce, M. J., 2011,
        ApJS 193, 12
   \bibitem[Ngeow \& Kanbur(2005)]{Ngeow2005} Ngeow, C-C., Kanbur, S. M., 2005,
        MNRAS 360, 1033
   \bibitem[Ngeow(2012)]{Ngeow2012} Ngeow, C-C., 2012,
        ApJ 747, 50
   \bibitem[Nordstr\"{o}m et al.(2004)]{Nord2004} Nordstr\"{o}m, B., Mayor, M., Andersen, J., Holmberg, J., Pont, F., J\o{}rgensen, B. R., Olsen, E. H., Udry, S., Mowlavi, N., 2004, 
	A\&A 418, 989
   \bibitem[Pancino et al.(2010)]{Pan2010} Pancino, E., Carrera, R., Rossetti, E., Gallart, C., 2010,
	A\&A 511, A56
   \bibitem[Pedicelli et al.(2009)]{Pedi2009} Pedicelli, S., Bono, G., Lemasle, B., Fran\c cois, P., Groenewegen, M., Lub, J., Pel, J. W., Laney, D., Piersimoni, A., Romaniello, M., Buonanno, R., Caputo, F., Cassisi, S., Castelli, F., Leurini, S., Pietrinferni, A., Primas, F., Pritchard, J., 2009,
        A\&A 504, 81
   \bibitem[Pedicelli et al.(2010)]{Pedi2010} Pedicelli, S., Lemasle, B., Groenewegen, M., Romaniello, M., Bono, G., Laney, D., Fran\c cois, P.,M., Buonanno, R., Caputo, F.,Lub, J., Pel, J. W., Primas, F., Pritchard, J., 2010, 
        A\&A 518, A11
   \bibitem[Perinotto \& Morbidelli(2006)]{Perr2006} Perinotto, M., Morbidelli, L., 2006, 
	MNRAS 372, 45
   \bibitem[Pilkington et al.(2012)]{Pil2012} 	Pilkington, K., Few, C. G., Gibson, B. K., Calura, F., Michel-Dansac, L., Thacker, R. J., Moll\' a, M., Matteucci, F., Rahimi, A., Kawata, D., Kobayashi, C., Brook, C. B., Stinson, G. S., Couchman, H. M. P., Bailin, J., Wadsley, J., 2012,
	A\&A 540, A56
   \bibitem[Piovan et al.(2011)]{Pio2012} Piovan, L., Chiosi, C., Merlin, E., Grassi, T., Tantalo, R., Buonomo, U., Cassar\'a, L. P., 2011, 		arXiv1107.4567P
   \bibitem[Portinari \& Chiosi(2000)]{Port2000} Portinari, L., Chiosi, C. 2000, 
	A\&A 355, 929
   \bibitem[Pottasch \& Bernard-Salas(2006)]{Pott2006} Pottasch, S. R., Bernard-Salas, J., 2006, 
	A\&A 457, 189 
   \bibitem[Prantzos \& Boissier(2000)]{Pran2000} Prantzos, N., Boissier, S., 2000,
	MNRAS 313, 338	
   \bibitem[Quillen et al.(2009)]{Quil2009} Quillen, A. C., Minchev, I., Bland-Hawthorn, J., Haywood M., 2009,
	MNRAS 397, 1599
   \bibitem[Quireza et al.(2006b)]{Qui2006b} Quireza, C., Rood, R. T., Bania, T. M., Balser, D. S., Maciel, W. J., 2006b
	ApJ 653, 1226
   \bibitem[Rahimi et al.(2011)]{Rahi2011} Rahimi, A., Kawata, D., Allende Prieto, C., Brook, C. B., Gibson, B. K., Kiessling, A., 2011,
	MNRAS 415, 1469
   \bibitem[Recio-Blanco et al.(2006)]{Recio2006} Recio-Blanco, A., Bijaoui, A., de Laverny, P., 2006,
	MNRAS 370, 141
   \bibitem[Reid et al.(2009)]{Reid2009} Reid, M. J., Menten, K. M., Zheng, X. W., Brunthaler, A., Xu, Y., 2009,
        ApJ 705, 1548
   \bibitem[Rolleston et al.(2000)]{Roll2000} Rolleston, W. R. J., Smartt, S. J., Dufton, P. L., Ryans, R. S. I., 2000, 
	A\&A, 363, 537
   \bibitem[Romaniello et al.(2008)]{Rom2008} Romaniello, M., Primas, F., Mottini, M., Pedicelli, S., Lemasle, B.,  Bono, G., Fran\c cois, P., Groenewegen, M. A. T., Laney, C. D., 2008, 
        A\&A 488, 731
   \bibitem[Ro\v{s}kar et al.(2008a)]{Ros2008a} Ro\v{s}kar, R., Debattista, V. P., Stinson, G. S., Quinn, T. R., Kaufmann, T., Wadsley, J., 2008a
	ApJ 675, 65	
   \bibitem[Ro\v{s}kar et al.(2008b)]{Ros2008b} Ro\v{s}kar, R., Debattista, V. P., Quinn, T. R., Stinson, G. S., Wadsley, J., 2008,
	ApJ 684, 79
   \bibitem[Rudolph et al.(2006)]{Rud2006} Rudolph, A. L., Fich, M., Bell, G. R., Norsen, T., Simpson, J. P., Haas, M. R., Erickson, E. F., 2006, 
	ApJS 162, 346
   \bibitem[Ryabchikova et al.(1999)]{Ryab1999} Ryabchikova, T. A., Piskunov, N. E., Stempels, H. C., Kupka, F., Weiss, W. W., 1999,
	Proc. of the 6th International Colloquium on Atomic Spectra and Oscillator Strengths, Victoria BC, Canada, 1998, Physica Scripta T83, 162
   \bibitem[Sanders et al.(2010)]{San2010} Sanders, N., Caldwell, N., McDowell, J., 2010,
        BAAS 36, 1113  
   \bibitem[Sasselov(1986)]{Sass1986} Sasselov, D. D., 1986, 
	PASP 98, 561
   \bibitem[Scarano \& L\' epine(2013)]{Sca2013} Scarano, S., L\' epine, J. R. D., 2013,
	MNRAS 428, 625
   \bibitem[Sch\" onrich \& Binney(2009)]{Schon2009} Sch\" onrich, R., Binney, J., 2009, 
	MNRAS 396, 203
   \bibitem[Sellwood \& Binney(2002)]{Sell2002} Sellwood, J. A., Binney, J. J., 2002, 
	MNRAS 336, 785
   \bibitem[Sestito et al.(2006)]{Ses2006} Sestito, P., Bragaglia, A., Randich, S., Carretta, E., Prisinzano, L., Tosi, M., 2006,
	A\&A 458, 121 
   \bibitem[Sestito et al.(2007)]{Ses2007} Sestito, P., Randich, S., Bragaglia, A., 2007,
	A\&A 465, 185
   \bibitem[Sestito et al.(2008)]{Ses2008} Sestito, P., Bragaglia, A., Randich, S., Pallavicini, R., Andrievsky, S. M., Korotin, S. A. 2008, 
	A\&A 488, 943
   \bibitem[Shappee et al.(2011)]{Sha2011} Shappee, B. J., Stanek, K. Z., 2011,
        ApJ 733, 124
   \bibitem[Simpson et al.(1995)]{Sim1995} Simpson, J. P., Colgan, S. W. J., Rubin, R. H., Erickson, E. F., Haas, M. R., 1995,
	ApJ 444, 721	
   \bibitem[Smartt \& Rolleston(1997)]{Sma1997} Smartt, S. J., Rolleston, W. R. J., 1997,
	ApJ 481, L47
   \bibitem[Soszy\'nski et al.(2005)]{Sos2005} Soszy\'nski, I., Gieren, W., Pietrzy\'nski, G., 2005, 
        PASP 117, 823
   \bibitem[Spitoni \& Matteucci(2011)]{Spi2011} Spitoni, E., Matteucci, F., 2011,
	A\&A 531, 72
   \bibitem[Spitzer \& Schwarzschild(1953)]{Spi1953} Spitzer, L., Schwarzschild, M., 1953
	ApJ 118, 106
   \bibitem[Stanghellini et al.(2006)]{Stan2006} Stanghellini, L., Guerrero, M. A., Cunha, K., Manchado, A., Villaver, E., 2006, 
	ApJ 651, 898
   \bibitem[Stanghellini \& Haywood(2010)]{Stan2010} Stanghellini, L., Haywood, M., 2010, 
	ApJ 714, 1096
   \bibitem[Steinmetz et al.(2006)]{Stein2006} Steinmetz, M., Zwitter, T., Siebert, A., Watson, F. G., Freeman, K. C., Munari, U., Campbell, R., Williams, M., Seabroke, G. M., Wyse, R. F. G. et al., 2006,
	AJ 132, 1645	
   \bibitem[Stinson et al.(2010)]{Stin2010} Stinson, G. S., Bailin, J., Couchman, H., Wadsley, J., Shen, S., Nickerson, S., Brook, C., Quinn, T., 2010,
	MNRAS 408, 812
   \bibitem[Storm et al.(2011a)]{Storm2011a} Storm, J., Gieren, W., Fouqu\'e, P., Barnes, T. G., Pietrzy\'nski, G., Nardetto, N., Weber, M., Granzer, T., Strassmeier, K. G., 2011a,
        A\&A 534, 94
   \bibitem[Storm et al.(2011b)]{Storm2011b} Storm, J., Gieren, W., Fouqu\'e, P., Barnes, T. G., Soszy\'nski, I., Pietrzy\'nski, G., Nardetto, N., Queloz, D., 2011b, 
        A\&A 534, 95
   \bibitem[Takeda et al.(2013)]{Take2013} Takeda, Y., Kang, D.-I., Han, I., Lee, B.-C., Kim, K.-M., 2013,
	MNRAS 432, 769	
   \bibitem[Tamman et al.(2003)]{Tamm2003} Tammann, G. A., Sandage, A., Reindl, B., 2003, 
        A\&A 404, 423
   \bibitem[Trippe et al.(2008)]{Trip2008} Trippe, S., Gillessen, S., Gerhard, O. E., Bartko, H., Fritz, T. K., Maness, H. L., Eisenhauer, F., Martins, F., Ott, T., Dodds-Eden, K., Genzel, R., 2008,
        A\&A 492, 419
   \bibitem[Twarog et al.(1997)]{Twa1997} Twarog, B. A., Ashman, K. M., Antony-Twarog, B. J., 1997, 
	AJ 114, 2556
   \bibitem[Usenko et al.(2011a)]{Us2011a} Usenko, I. A., Kniazev, A. Y., Berdnikov, L. N., Kravtsov, V. V., 2011,
	AstL 37, 499
\bibitem[Usenko et al.(2011b)]{Us2011b} Usenko, I. A., Berdnikov, L. N., Kravtsov, V. V., Kniazev, A. Y., Chini, R., Hoffmeister, V. H., Stahl, O., Drass, H., 2011,
	AstL 37, 718
   \bibitem[Vilchez \&  Esteban(1996)]{Vil1996} Vilchez, J. M., Esteban, C., 1996, 
	MNRAS 280, 720
   \bibitem[Vlaji\'c et al.(2009)]{Vla2009} Vlaji\'c, M., Bland-Hawthorn, J., Freeman, K. C., 2009,
        ApJ 697, 361
   \bibitem[Vlaji\'c et al.(2009)]{Vla2011} Vlaji\'c, M., Bland-Hawthorn, J., Freeman, K. C., 2011,
        ApJ 732, 7
   \bibitem[Wang \& Zhao(2013)]{Wang2013} Wang, Y., Zhao, G., 2013,
	ApJ 769, 4
   \bibitem[Wiersma et al.(2011)]{Wier2011} Wiersma, R. P. C., Schaye, J., Theuns, T.. 2011, 
	MNRAS 415, 353
   \bibitem[Yanni et al.(2009)]{Yanni2009} 	Yanny, B., Rockosi, C., Newberg, H. J., Knapp, G. R. et al, 2009
	AJ 137, 4377
   \bibitem[Yong et al.(2005)]{Yong2005} Yong, D., Carney, B. W., de Almeida, M. L. T., 2005, 
	AJ 130, 597
   \bibitem[Yong et al.(2006)]{Yong2006} Yong, D., Carney, B. W., de Almeida, M. L. T., Pohl, B. L., 2006, 
	AJ 131, 2256
   \bibitem[Yong et al.(2012)]{Yong2012} Yong, D., Carney, B. W., Friel, E. D., 2012,
	AJ 144, 95	
\end{thebibliography}
\end{document}